\documentclass[11pt,english]{article}
\usepackage[]{graphicx}
\usepackage{amsmath,amsfonts,amssymb}

\def\subsvar{\delta_{(s)}}

\begin{document}

\title{Multipolar test body equations of motion in generalized gravity theories}

\author{Yuri N. Obukhov\footnote{Email: obukhov@ibrae.ac.ru} \\
        Theoretical Physics Laboratory\\
				Nuclear Safety Institute\\
        Russian Academy of Sciences\\
				B.Tulskaya 52, 115191 Moscow, Russia \\
				\\
Dirk Puetzfeld\footnote{Email: dirk.puetzfeld@zarm.uni-bremen.de, URL: http://puetzfeld.org} \\
        ZARM\\
        University of Bremen\\
        Am Fallturm, 28359 Bremen, Germany}

\date{\today}

\maketitle

\begin{abstract}
We give an overview of the derivation of multipolar equations of motion of extended test bodies for a wide set of gravitational theories beyond the standard general relativistic framework. The classes of theories covered range from simple generalizations of General Relativity, e.g.\ encompassing additional scalar fields, to theories with additional geometrical structures which are needed for the description of microstructured matter. Our unified framework even allows to handle theories with nonminimal coupling to matter, and thereby for a systematic test of a very broad range of gravitational theories.
\end{abstract}

\section{Introduction}\label{introduction_sec}

In this work we present a general unified multipolar framework, which enables us to derive equations of motion of extended test bodies for a wide range of gravitational theories. The framework presented here can be applied to theories which significantly go beyond General Relativity (GR), and range from the most straightforward extensions of GR, like scalar-tensor theories, to theories with additional geometrical structures and nonminimal coupling. 

\begin{figure}
\begin{center}
\includegraphics[height=6cm]{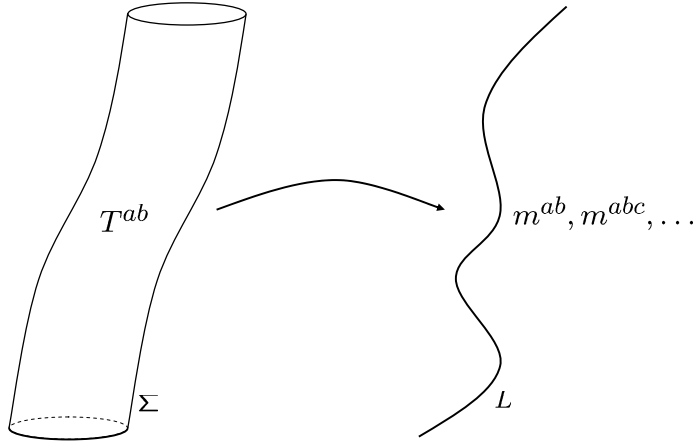}
\end{center}
\caption{General idea behind multipolar approximation schemes: The world-tube $\Sigma$ of a body is replaced by a representative world-line $L$, whereas the original energy-momentum tensor $T^{ab}$ is substituted by a set of multipole moments $m^{ab \cdots}$ along this world-line. Such a multipolar description simplifies the equations of motion. This is achieved by consideration of only a finite set of moments. Different flavors of multipolar approximation schemes exist in the literature, in this work we define the moments  \`{a} la Dixon in \cite{Dixon:1964}.}
\label{sample_fig:1}
\end{figure}

The {\it multipolar method} which we employ here can be thought of as the direct generalization of the ideas  pioneered by Mathisson \cite{Mathisson:1937}, Papapetrou \cite{Papapetrou:1951:3}, and Dixon \cite{Dixon:1964,Dixon:1974,Dixon:1979,Dixon:2008} to the case of generalized gravity theories. As sketched in figure \ref{sample_fig:1}, the main aim of such methods is to find a simplified {\it local} description of the motion of extended test bodies in terms of a suitable set of multipolar moments, which catches the essential properties of the body at the chosen order of approximation.

In this work, we use a covariant multipolar description, based on Synge's world-function formalism \cite{Synge:1960,DeWitt:Brehme:1960}. The {\it multipolar moments} to be introduced, can be viewed as a direct generalization of the moments first introduced by Dixon in \cite{Dixon:1964}. 

Central to the derivation of the equations of motion, by means of a multipolar method, is the knowledge of the corresponding conservation laws of the underlying gravity theory. In General Relativity, the starting point of all methods mentioned so-far is the conservation of the (symmetric) energy-momentum 
\begin{equation}
\nabla_j T^{ij} = 0. \label{em_conservation_gr}
\end{equation}
Many generalized gravity theories allow for a more detailed description of matter than standard GR, e.g.\ also taking into account internal matter degrees of freedom as source of the gravitational field. Consequently the conservation laws of such theories are more complex, and (\ref{em_conservation_gr}) has to be replaced by a suitable generalization.

\paragraph{Generalized gravity}

A metric and connection are the two fundamental geometrical objects on a spacetime manifold. They play an important role in the description of gravitational phenomena in the framework of what can be quite generally called an Einsteinian approach to gravity. The principles of equivalence and general coordinate covariance are the cornerstones of this approach. As Einstein himself formulated, the crucial achievement of his theory was the elimination of the notion of inertial systems as preferred ones among all possible coordinate systems. 

In Einstein's theory, gravitation is associated with the metric tensor alone. Nevertheless, it is worthwhile to stress that Einstein clearly understood the different physical statuses of the metric and the connection (Appendix II in \cite{Einstein:1956}): 

\begin{quotation}
``[...] at first Riemannian metric was considered the fundamental concept on which the general theory of relativity and thus the avoidance of the inertial system were based. Later, however, Levi-Civita rightly pointed out that the element of the theory that makes it possible to avoid the inertial system is rather the infinitesimal displacement field $\Gamma_{ik}{}^j$. The metric or the symmetric tensor field $g_{ik}$ which defines it is only indirectly connected with the avoidance of the inertial system insofar as it determines a displacement field.''.
\end{quotation}

There exists a variety of gravitational theories that generalize or extend the physical and mathematical structure of GR. Among these theories there are large classes of so-called $f(R)$ models, and of theories with nonminimal coupling to matter; they are developed in particular in the context of relativistic cosmology (but not only there), see \cite{Harko:2014:1,Schmidt:2007,Bertolami:etal:2007,Straumann:2008,Nojiri:2011}. The so-called Palatini approach represents another class of widely discussed theories in which the metric and the connection are treated as independent variables in the action principle \cite{Hehl:1978,Hehl:1981,Sotiriou:2010}. Another early example of a generalized gravity theory with balance laws different from Einstein's gravity can be found in \cite{Rastall:1972,Smalley:1974}. Last but not least, we should mention the vast family of the gauge gravity theories constructed using a Yang-Mills type of approach \cite{Blagojevic:2002,Hehl:2013}. 

\begin{figure}
\begin{center}
\includegraphics[width=8cm, angle=-90]{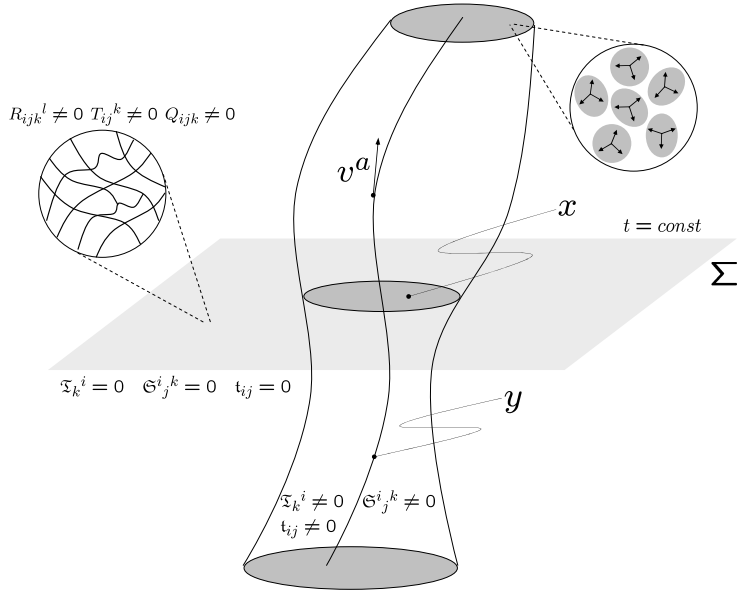}
\end{center}
\caption{In contrast to GR, MAG also allows to take into account the microstructural properties of matter (spin, dilation current, proper hypercharge). These additional currents couple to the post-Riemannian geometric degrees of freedom of the underlying spacetime. Our unified description of extended test bodies allows to cover a wide range of gravitational theories and matter models.}
\label{fig_world_tube}
\end{figure}

Since our main aim is to present a unified multipolar framework for a very wide range of gravity theories, the theory underlying our analysis has to be sufficiently general. In this work we choose metric-affine gravity (MAG) as the foundation for the derivation of the equations of motion of extended deformable test bodies.

In MAG, matter is characterized by three fundamental Noether currents -- the canonical energy-momentum current, the canonical hypermomentum current, and the metrical energy-momentum current. These objects satisfy a set of conservation laws (or, more exactly, balance equations). In view of the multi-current characterization of matter in metric-affine gravity, we develop a general approach which is applicable to an arbitrary set of conservation laws for any number of currents. The latter can include gravitational, electromagnetic, and other physical currents if they are relevant to the model under consideration. In particular, our approach is flexible enough to be applied to the case in which there is a general nonminimal coupling between gravity to matter. Our presentation here is mainly based on the original works \cite{Obukhov:Puetzfeld:2014,Obukhov:Puetzfeld:2014:2,Puetzfeld:Obukhov:2014}. 

\paragraph{Structure of the paper}

The structure of the paper is as follows: In section \ref{sec_models} we give an overview of the metric-affine theory of gravity. In particular we cover its geometrical and dynamical aspects. This is followed by a general Lagrange-Noether analysis in section \ref{Noether_sec}. The results of this general analysis are then used to derive the conservation laws of a metric-affine gravity with nonminimal coupling in section \ref{conservation_sec}. In section \ref{master_sec} a unified multipolar framework is presented. This framework is then used \ref{eom_sec} to derive the general equations of motion of extended test bodies in nonminimal metric-affine gravity. Several specializations to other gravity theories are discussed. Furthermore, in section \ref{scalar_tensor_eom_sec} the equations of motion in scalar-tensor theories are worked out. Section \ref{conclusion_sec} contains a summary of our results.

Our conventions are those of \cite{Hehl:1995}. In particular, the basic geometrical quantities such as the curvature, torsion, and nonmetricity are defined as in \cite{Hehl:1995}, and we use the Latin alphabet to label the spacetime coordinate indices. Furthermore, the metric has the signature $(+,-,-,-)$. It should be noted that our definition of the metrical energy-momentum tensor differs by a sign from the definition used in \cite{Bertolami:etal:2007,Nojiri:2011,Puetzfeld:Obukhov:2013}. A summary of our notation and conventions, as well as additional material for comparison to previous works can be found in the appendix. 

\section{Metric-affine gravity primer} \label{sec_models}

Metric-affine gravity \cite{Hehl:1995} is a natural extension of Einstein's general relativity theory. It is based on gauge-theoretic principles \cite{Blagojevic:2002,Hehl:2013}, and it takes into account microstructural properties of matter (spin, dilation current, proper hypercharge) as possible physical sources of the gravitational field on an equal footing with macroscopic properties (energy and momentum) of matter. The formalism of MAG makes it possible to study all the above-mentioned alternative theories in a unified framework. The corresponding spacetime landscape \cite{Schouten:1954} includes as special cases the geometries of Riemann, Riemann-Cartan, Weyl, Weitzenb\"ock, etc.\ (c.f.\ figure \ref{fig:1}). 

The standard understanding of metric-affine gravity is based on the gauge-theoretic approach for the general affine symmetry group. In this framework, we can naturally distinguish the {\it kinematics} of the gravity theory and its {\it dynamics}. The kinematics embraces all aspects that are related to the description of the fundamental variables, their mathematical properties and physical interpretation. The dynamics studies the choice of the Lagrangian and the field equations. We collect in this section all the relevant material. Although the Lagrange-Noether machinery is deeply interrelated with both kinematics and dynamics of the theory, we will discuss it in a separate section which will ultimately underlie the derivation of the multipole equations of motion. 

\subsection{Kinematics of metric-affine gravity}

We model spacetime as a four-dimensional smooth manifold. In this study we are not interested in the global (topological) aspects, and we will confine our attention only to local issues. The local coordinates $x^i$, $i = 0,1,2,3$, are introduced in the neighborhood of an arbitrary point of the spacetime manifold. The geometrical (gravitational) and physical (material) variables are then fields of different nature (both tensors and nontensors) over the spacetime. They are characterized by their components and transformation properties under local diffeomorphisms $x^i\rightarrow x^i + \delta x^i$, where
\begin{equation}
\delta x^i = \xi^i(x).\label{dex}
\end{equation}
The four arbitrary functions $\xi^i(x)$ parametrize an arbitrary local diffeomorphism. 

\subsubsection{Geometrical variables}

In the metric-affine theory of gravity, the gravitational physics is encoded in two fields: the metric tensor $g_{ij}$ and an independent linear connection $\Gamma_{ki}{}^j$. The latter is not necessarily symmetric and compatible with the metric. From the geometrical point of view, the metric introduces lengths and angles of vectors, and thereby determines the distances (intervals) between points on the spacetime manifold. The connection introduces the notion of parallel transport and defines the covariant differentiation $\nabla_k$ of tensor fields. 

Under the spacetime diffeomorphisms (\ref{dex}), these geometrical variables transform as 
\begin{eqnarray}
\delta g_{ij} &=& -\,(\partial_i\xi^k)\,g_{kj} - (\partial_j\xi^k)\,g_{ik},\label{dgij}\\
\delta\Gamma_{ki}{}^j &=&  -\,(\partial_k\xi^l)\,\Gamma_{li}{}^j - (\partial_i\xi^l)\,\Gamma_{kl}{}^j + (\partial_l\xi^j)\,\Gamma_{ki}{}^l - \partial^2_{ki}\xi^j.\label{dG}
\end{eqnarray}

In general, the geometry of a metric-affine manifold is exhaustively characterized by three tensors: the curvature, the torsion and the nonmetricity. They are defined \cite{Schouten:1954} as follows:
\begin{eqnarray}
R_{kli}{}^j &:=& \partial_k\Gamma_{li}{}^j - \partial_l\Gamma_{ki}{}^j + \Gamma_{kn}{}^j \Gamma_{li}{}^n - \Gamma_{ln}{}^j\Gamma_{ki}{}^n,\label{curv}\\
T_{kl}{}^i &:=& \Gamma_{kl}{}^i - \Gamma_{lk}{}^i,\label{tors}\\ \label{nonmet}
Q_{kij} &:=& -\,\nabla_kg_{ij} = - \partial_kg_{ij} + \Gamma_{ki}{}^lg_{lj} + \Gamma_{kj}{}^lg_{il}.
\end{eqnarray}
The curvature and the torsion tensors determine the commutator of the covariant derivatives. For a tensor $A^{c_1 \dots c_k}{}_{d_1 \dots d_l}$ of arbitrary rank and index structure: 
\begin{eqnarray}
&& (\nabla_a\nabla_b - \nabla_b\nabla_a) A^{c_1 \dots c_k}{}_{d_1 \dots d_l} = -\,T_{ab}{}^e\nabla_e A^{c_1 \dots c_k}{}_{d_1 \dots d_l} \nonumber \\
&& + \sum^{k}_{i=1} R_{abe}{}^{c_i} A^{c_1 \dots e \dots c_k}{}_{d_1 \dots d_l} - \sum^{l}_{j=1}R_{abd_j}{}^{e} A^{c_1 \dots c_k}{}_{d_1 \dots e \dots d_l}. \label{commutator}
\end{eqnarray}
The Ricci tensor is introduced by $R_{ij} := R_{kij}{}^k$, and the curvature scalar is $R := g^{ij}R_{ij}$. 

\begin{figure}
\begin{center}
\includegraphics[height=6cm]{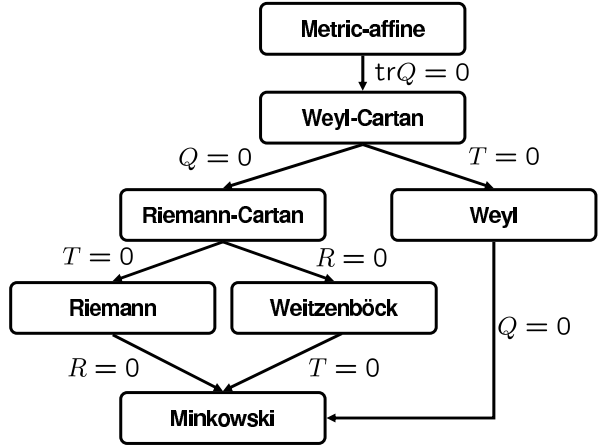}
\end{center}
\caption{Different spacetime types as special cases of a general metric-affine spacetime ($R_{kli}{}^j \neq 0$, $T_{kl}{}^i \neq 0$, $Q_{kij} \neq 0$). The abbreviations $R$ (curvature), $T$ (torsion), and $Q$ (nonmetricity) over the arrows denote the vanishing of the corresponding geometrical object.}
\label{fig:1}
\end{figure}

A general metric-affine spacetime ($R_{kli}{}^j \neq 0$, $T_{kl}{}^i \neq 0$, $Q_{kij} \neq 0$) incorporates several other spacetimes as special cases, see figure \ref{fig:1} for an overview.
The Riemannian connection $\widetilde{\Gamma}_{kj}{}^i$ is uniquely determined by the conditions of vanishing torsion and nonmetricity which yield explicitly 
\begin{equation}
\widetilde{\Gamma}_{kj}{}^i = {\frac 12}g^{il}(\partial_jg_{kl} + \partial_kg_{lj} - \partial_lg_{kj}).\label{Chr}
\end{equation}
Here and in the following, a tilde over a symbol denotes a Riemannian object (such as the curvature tensor) or a Riemannian operator (such as the covariant derivative) constructed from the Christoffel symbols (\ref{Chr}). The deviation of the geometry from the Riemannian one is then conveniently described by the {\it distortion} tensor 
\begin{equation}
N_{kj}{}^i := \widetilde{\Gamma}_{kj}{}^i - \Gamma_{kj}{}^i.\label{dist}
\end{equation}
The system (\ref{tors}) and (\ref{nonmet}) allows to find the distortion tensor in terms of the torsion and nonmetricity. Explicitly:
\begin{equation}
N_{kj}{}^i = -\,{\frac 12}(T_{kj}{}^i + T^i{}_{kj} + T^i{}_{jk})  + {\frac 12}(Q^i{}_{kj} - Q_{kj}{}^i - Q_{jk}{}^i).\label{NTQ}
\end{equation}
Conversely, one can use this to express the torsion and nonmetricity tensors in terms of the distortion,
\begin{eqnarray}
T_{kj}{}^i &=& -\,2N_{[kj]}{}^i,\label{TN}\\
Q_{kij} &=& -\,2N_{k(ij)}.\label{QN}
\end{eqnarray}
Substituting (\ref{dist}) into (\ref{curv}), we find the relation between the non-Riemannian and the Riemannian curvature tensors
\begin{equation}
R_{adc}{}^b = \widetilde{R}_{adc}{}^b - \widetilde{\nabla}_aN_{dc}{}^b + \widetilde{\nabla}_dN_{ac}{}^b + N_{an}{}^bN_{dc}{}^n - N_{dn}{}^bN_{ac}{}^n.\label{RRN}
\end{equation}

Applying the covariant derivative to (\ref{curv})-(\ref{nonmet}) and antisymmetrizing, we derive the Bianchi identities \cite{Schouten:1954}:
\begin{eqnarray}
\nabla_{[n}R_{kl]i}{}^j &=& T_{[kl}{}^m R_{n]mi}{}^j,\label{Dcurv}\\
\nabla_{[n}T_{kl]}{}^i &=& R_{[kln]}{}^i + T_{[kl}{}^m T_{n]m}{}^i,\label{Dtors}\\ 
\nabla_{[n}Q_{k]ij} &=& R_{nk(ij)}.\label{Dnonmet}
\end{eqnarray}

\subsubsection{Tensors, densities, and covariant differential operators} 

Along with tensors, an important role in physics is played by densities. A fundamental density $\sqrt{-g}$ is constructed from the determinant of the metric, $g=$det$g_{ij}$. Under diffeomorphisms (\ref{dex}) it transforms as
\begin{equation}
\delta\sqrt{-g} = - \,(\partial_i\xi^i)\,\sqrt{-g}.\label{ddet}
\end{equation}
This is a direct consequence of (\ref{dgij}). From any tensor $B_{i\dots}{}^{j\dots}$ one can construct a density ${\mathfrak B}_{i\dots}{}^{j\dots} = \sqrt{-g}B_{i\dots}{}^{j\dots}$. Although in this paper we will encounter only such objects, it is worthwhile to notice that not all densities are of this type -- see the exhaustive presentation in the book of Synge and Schild \cite{Synge:1978}. 

There are two kinds of covariant differential operators on the spacetime manifold, depending on whether the connection is involved or not. The Lie derivative ${\cal L}_\zeta$ is defined along any vector field $\zeta^i$ and it maps tensors (densities) into tensors (densities) of the same rank. Let us recall the explicit form of the Lie derivative of the metric and the distortion:
\begin{eqnarray}
{\cal L}_\zeta g_{ij} &=& \zeta^k\partial_kg_{ij} + (\partial_i\zeta^k)g_{kj} + (\partial_j\zeta^k)g_{ik},\label{Lm}\\
{\cal L}_\zeta N_{kj}{}^i &=& \zeta^n\partial_nN_{kj}{}^i + (\partial_k\zeta^n)N_{nj}{}^i + (\partial_j\zeta^n)N_{kn}{}^i - (\partial_n\zeta^i)N_{kj}{}^n.\label{Ld}
\end{eqnarray}

In contrast, a covariant derivative $\nabla_k$ raises the rank of tensors (densities) and it is determined by the linear connection $\Gamma_{kj}{}^i$. Moreover, there are different covariant derivatives which arise for different connections that may coexist on the same manifold. 

A mathematical fact is helpful in this respect: Every third rank tensor $X_{kj}{}^i$ defines a map of one connection into a different new connection 
\begin{equation}
\Gamma_{kj}{}^i\quad\longrightarrow\quad\Gamma_{kj}{}^i + X_{kj}{}^i.\label{G2G}
\end{equation} 
There are important special cases of such a map. One example is obtained for $X_{kj}{}^i = N_{kj}{}^i$: then the connection $\Gamma_{kj}{}^i$ is mapped into the Riemannian Christoffel symbols, $\widetilde{\Gamma}_{kj}{}^i = \Gamma_{kj}{}^i + N_{kj}{}^i$, in accordance with (\ref{dist}). 

Another interesting case arises for $X_{kj}{}^i = T_{jk}{}^i$. The result of the mapping
\begin{equation}
\overline{\Gamma}_{kj}{}^i = \Gamma_{kj}{}^i + T_{jk}{}^i = \Gamma_{kj}{}^i + \Gamma_{jk}{}^i - \Gamma_{kj}{}^i = \Gamma_{jk}{}^i.\label{tr}
\end{equation}
is then called a {\it transposed connection}, or associated connection, see \cite{Lichnerowicz:1977,Trautman:1973}. 

The importance of the transposed connection is manifest in the following observation. Although the Lie derivative is a covariant operator -- this is not apparent since it is based on partial derivatives -- one can make everything explicitly covariant by noticing that it is possible to recast (\ref{Lm}) and (\ref{Ld}) into equivalent form
\begin{eqnarray}
{\cal L}_\zeta g_{ij} &=& \zeta^k\nabla_kg_{ij} + (\overline{\nabla}_i\zeta^k)g_{kj} + (\overline{\nabla}_j\zeta^k)g_{ik},\label{LmC}\\
{\cal L}_\zeta N_{kj}{}^i &=& \zeta^n\nabla_nN_{kj}{}^i + (\overline{\nabla}_k\zeta^n)N_{nj}{}^i + (\overline{\nabla}_j\zeta^n)N_{kn}{}^i - (\overline{\nabla}_n\zeta^i)N_{kj}{}^n.\label{LdC}
\end{eqnarray}
By the same token we can ``covariantize'' the Lie derivatives for all other tensors of any structure and of arbitrary rank. 

A more nontrivial (and less known) fact is that we can define the Lie derivatives also for objects which are not tensors. In particular, the Lie derivative of the connection then reads \cite{Lichnerowicz:1977}:
\begin{equation}\label{LieGam}
{\cal L}_\zeta\Gamma_{kj}{}^i = \nabla_k\overline{\nabla}_j\zeta^i - R_{klj}{}^i\zeta^l.
\end{equation}
This quantity measures the noncommutativity of the Lie derivative with the covariant derivative
\begin{eqnarray}
&&({\cal L}_\zeta \nabla_k - \nabla_k{\cal L}_\zeta) A^{c_1 \dots c_k}{}_{d_1 \dots d_l} \nonumber\\
&& = \,\sum^{k}_{i=1} ({\cal L}_\zeta\Gamma_{kb}{}^{c_i}) A^{c_1 \dots b \dots c_k}{}_{d_1 \dots d_l} - \sum^{l}_{j-1} ({\cal L}_\zeta\Gamma_{kd_j}{}^{b}) A^{c_1 \dots c_k}{}_{d_1 \dots b \dots d_l}. \label{commLD}
\end{eqnarray}

The connection $\Gamma_{kj}{}^i$, the transposed connection  $\overline{\Gamma}_{kj}{}^i$, and the Riemannian connection  $\widetilde{\Gamma}_{kj}{}^i$ define the three respective covariant derivatives: $\nabla_k$, $\overline{\nabla}_k$, and $\widetilde{\nabla}_k$. The covariant derivative defined by the Riemannian connection is conventionally denoted by a semicolon ``$ {}_{;a}$''. 

We will assume that these differential operators act on tensors. In addition, we will need the covariant operators that act on densities. For an arbitrary tensor density ${\mathfrak B}^n{}_{i\dots}{}^{j\dots}$ we introduce the covariant derivative 
\begin{equation}
\widehat{\nabla}{}_n{\mathfrak B}^n{}_{i\dots}{}^{j\dots} := \partial_n{\mathfrak B}^n{}_{i\dots}{}^{j\dots} + \Gamma_{nl}{}^j{\mathfrak B}^n{}_{i\dots}{}^{l\dots} - \Gamma_{ni}{}^l {\mathfrak B}^n{}_{l\dots}{}^{j\dots},\label{dA}
\end{equation}
which produces a tensor density of the same weight. We denote a similar differential operation constructed with the help of the Riemannian connection by
\begin{equation}
\check{\nabla}{}_n{\mathfrak B}^n{}_{i\dots}{}^{j\dots}:= \partial_n{\mathfrak B}^n{}_{i\dots}{}^{j\dots} + \widetilde{\Gamma}_{nl}{}^j{\mathfrak B}^n{}_{i\dots}{}^{l\dots} - \widetilde{\Gamma}_{ni}{}^l{\mathfrak B}^n{}_{l\dots}{}^{j\dots},\label{dAc}
\end{equation}
When ${\mathfrak B}^n{}_{i\dots}{}^{j\dots} = \sqrt{-g}B^n{}_{i\dots}{}^{j\dots}$, we find 
\begin{eqnarray}
\widehat{\nabla}_n{\mathfrak B}^n{}_{i\dots}{}^{j\dots} &=& \sqrt{-g}\,{\stackrel * \nabla}{}_nB^n{}_{i\dots}{}^{j\dots},\label{nablastar}\\
\check{\nabla}_n{\mathfrak B}^n{}_{i\dots}{}^{j\dots} &=& \sqrt{-g}\,\widetilde{\nabla}{}_nB^n{}_{i\dots}{}^{j\dots},\label{nablatilde}
\end{eqnarray}
where we introduced a modified covariant derivative
\begin{equation}
{\stackrel * \nabla}{}_i := \nabla_i + N_{ki}{}^k.\label{dstar}
\end{equation}

\subsubsection{Matter variables}

We will not specialize the discussion of matter to any particular physical field. It will be more convenient to describe matter by a generalized field $\psi^A$. The range of the indices $A,B,\dots$ is not important in our study. However, we do need to know the behavior of the matter field under spacetime diffeomorphisms (\ref{dex}):
\begin{equation}
\delta\psi^A = -\,(\partial_i\xi^j)\,(\sigma^A{}_B)_j{}^i\,\psi^B.\label{dpsiA}
\end{equation}
Here $(\sigma^A{}_B)_j{}^i$ are the generators of general coordinate transformations that satisfy the commutation relations
\begin{eqnarray}
(\sigma^A{}_C)_j{}^i(\sigma^C{}_B)_l{}^k - (\sigma^A{}_C)_l{}^k (\sigma^C{}_B)_j{}^i
= (\sigma^A{}_B)_l{}^i\,\delta^k_j - (\sigma^A{}_B)_j{}^k \,\delta^i_l.\label{comms}
\end{eqnarray}
We immediately recognize in (\ref{comms}) the Lie algebra of the general linear group $GL(4,R)$. This fact is closely related to the standard gauge-theoretic interpretation \cite{Hehl:1995} of metric-affine gravity as the gauge theory of the general affine group $GA(4,R)$, which is a semidirect product of spacetime translation group times $GL(4,R)$.

The transformation properties (\ref{dpsiA}) determine the form of the covariant and the Lie derivative of a matter field: 
\begin{eqnarray}\label{Dpsi}
\nabla_k\psi^A &:=& \partial_k\psi^A -\Gamma_{ki}{}^j\,(\sigma^A{}_B)_j{}^i\,\psi^B,\\ \label{Lpsi}
{\cal L}_\zeta\psi^A &:=& \zeta^k\partial_k\psi^A + (\partial_i\zeta^j)(\sigma^A{}_B)_j{}^i\,\psi^B.
\end{eqnarray}
The commutators of these differential operators read
\begin{eqnarray}
(\nabla_k\nabla_l - \nabla_l\nabla_k)\psi^A &=& -\,R_{klj}{}^i(\sigma^A{}_B)_i{}^j\psi^B - T_{kl}{}^i\nabla_i\psi^A,\label{comDDpsi}\\
 ({\cal L}_\zeta \nabla_k - \nabla_k{\cal L}_\zeta)\psi^A &=& -\,({\cal L}_\zeta\Gamma_{kj}{}^i)(\sigma^A{}_B)_i{}^j\psi^B.\label{comDLpsi}
\end{eqnarray}

\subsubsection{Symmetries in MAG: generalized Killing vectors}\label{killing_sec}

As is well known, symmetries of a Riemannian spacetime are generated by Killing vector fields. Each such field defines a so-called {\it motion} of the spacetime manifold, that is a diffeomorphism which preserves the metric $g_{ij}$. 

Suppose $\zeta^i$ is a Killing vector field. By definition, it satisfies
\begin{equation}
\widetilde{\nabla}_i\zeta_j + \widetilde{\nabla}_j\zeta_i = 0.\label{K1}
\end{equation}
By differentiation, we derive from this the second covariant derivative 
\begin{equation}
\widetilde{\nabla}_i\widetilde{\nabla}_j\zeta_k = \widetilde{R}_{jki}{}^l\zeta_l.\label{K2}
\end{equation}
Apply another covariant derivative and antisymmetrize: 
\begin{equation}
\widetilde{\nabla}_{[n}\widetilde{\nabla}_{i]}\widetilde{\nabla}_j\zeta_k = \widetilde{\nabla}_{[n}(\widetilde{R}_{|jk|i]}{}^l\zeta_l).\label{K2a}
\end{equation}
After some algebra, the last equation is recast into
\begin{eqnarray}
\zeta^n\widetilde{\nabla}_n\widetilde{R}_{ijkl} + \widetilde{R}_{njkl}\widetilde{\nabla}_i\zeta^n + \widetilde{R}_{inkl}\widetilde{\nabla}_j\zeta^n + \widetilde{R}_{ijnl}\widetilde{\nabla}_k\zeta^n + \widetilde{R}_{ijkn}\widetilde{\nabla}_l\zeta^n = 0.\label{K3}
\end{eqnarray}

The equations (\ref{K1}), (\ref{K2}) and (\ref{K3}) have a geometrical meaning:
\begin{eqnarray}
{\cal L}_\zeta g_{ij} &=& 0,\label{Lg}\\
{\cal L}_\zeta \widetilde{\Gamma}_{ij}{}^k &=& 0,\label{LG}\\
{\cal L}_\zeta \widetilde{R}_{ijkl} &=& 0.\label{LR}
\end{eqnarray}
That is, the Lie derivatives along the Killing vector field $\zeta$ vanish for all Riemannian geometrical objects. Moreover, one can show that the same is true for all higher covariant derivatives of the Riemannian curvature tensor \cite{Yano:1955}
\begin{equation}
{\cal L}_\zeta \left(\widetilde{\nabla}_{n_1}\dots \widetilde{\nabla}_{n_N}\widetilde{R}_{ijkl}\right) = 0.\label{LDR} 
\end{equation}

Let us generalize the notion of a symmetry to the metric-affine spacetime. We take an ordinary Killing vector field $\zeta$ and postulate the vanishing of the Lie derivative 
\begin{equation}
{\cal L}_\zeta N_{kj}{}^i = 0\label{LieN}
\end{equation}
of the distortion tensor. Combining this with (\ref{NTQ}) and (\ref{LG}), we find an equivalent formulation
\begin{eqnarray}
{\cal L}_\zeta g_{ij} &=& 0,\label{LgMAG}\\
{\cal L}_\zeta \Gamma_{ij}{}^k &=& 0.\label{LGMAG}
\end{eqnarray}
We call a vector field that satisfies (\ref{LgMAG}) and (\ref{LGMAG}) a {\it generalized Killing vector} of the metric-affine spacetime. By definition, such a $\zeta$ generates a diffeomorphism of the spacetime manifold that is simultaneously an isometry (\ref{LgMAG}) and an isoparallelism (\ref{LGMAG}). 

Since the Lie derivative along a Killing vector commutes with the covariant derivative, ${\cal L}_\zeta\widetilde{\nabla}_i = \widetilde{\nabla}_i {\cal L}_\zeta$, see (\ref{commLD}), we conclude from (\ref{RRN}) and (\ref{LR}) that the generalized Killing vector leaves the non-Riemannian curvature tensor invariant
\begin{equation}
{\cal L}_\zeta R_{klj}{}^i = 0.\label{LRMAG}
\end{equation}
It is also straightforward to verify that 
\begin{equation}
{\cal L}_\zeta \left(\nabla_{n_1}\dots \nabla_{n_N}R_{klj}{}^i\right) = 0\label{LDRMAG} 
\end{equation}
for any number of covariant derivatives of the curvature.

Later we will show that generalized Killing vectors have an important property: they induce conserved quantities on the metric-affine spacetime. 

\subsection{Dynamics of metric-affine gravity}\label{fieldeqs_sec}

The explicit form of the dynamical equations of the gravitational field is irrelevant for the conservation laws that will form the basis for the derivation of the test body equations of motion. However, for completeness, we discuss here the field equations of a general metric-affine theory of gravity. The standard understanding of MAG is its interpretation as a gauge theory based on the general affine group $GA(4,R)$, which is a semidirect product of the general linear group $GL(4,R)$, and the group of local translations \cite{Hehl:1995}. The corresponding gauge-theoretic formalism generalizes the approach of Sciama and Kibble \cite{Sciama:1962,Kibble:1961}; for more details about gauge gravity theories, see \cite{Blagojevic:2002,Hehl:2013}. In the standard formulation of MAG as a gauge theory \cite{Hehl:1995}, the gravitational gauge potentials are identified with the metric, coframe, and the linear connection. The corresponding gravitational field strengths are then the nonmetricity, the torsion, and the curvature, respectively. 

Here we use an alternative formulation of MAG, in which gravity is described by a different set of fundamental field variables, i.e.\ the independent metric $g_{ij}$ and connection $\Gamma_{ki}{}^j$, see \cite{Hehl:1976a,Hehl:1976b,Hehl:1976c,Hehl:1978,Hehl:1981,Obukhov:1982,Vassiliev:2005,Sotiriou:2007,Sotiriou:2010,Vitagliano:2011}, for example. It is worthwhile to compare the field equations in the different formulations of MAG, and in particular, it is necessary to clarify the role and place of the {\it canonical} energy-momentum tensor as a source of the gravitational field. Since one does not have the coframe (tetrad) among the fundamental variables, the corresponding field equation is absent. Here we demonstrate that one can always rearrange the field equations of MAG in such a way that the canonical energy-momentum tensor is recovered as one of the sources of the gravitational field. 

Assuming standard minimal coupling, the total Lagrangian density of interacting gravitational and matter fields reads
\begin{equation}\label{Ltot}
{\mathfrak L} = {\mathfrak V}(g_{ij}, R_{ijk}{}^l, N_{ki}{}^j) + {\mathfrak L}_{\rm mat}(g_{ij}, \psi^A, \nabla_i\psi^A).
\end{equation}
In general, the gravitational Lagrangian density ${\mathfrak V}$ is constructed as a diffeomorphism invariant function of the curvature, torsion, and nonmetricity. However, in view of the relations (\ref{TN}) and (\ref{QN}), we can limit ourselves to Lagrangian functions that depend arbitrarily on the curvature and the distortion tensors. The matter Lagrangian depends on the matter fields $\psi^A$ and their {\it covariant derivatives} (\ref{Dpsi}).

The field equations of metric-affine gravity can be written in several equivalent ways. The standard form is the set of the so-called ``first'' and ``second'' field equations. Using the {\it modified covariant derivative} defined by (\ref{dstar}), the field equations are given by
\begin{eqnarray}
\check{\nabla}{}_n {\mathfrak H}^{in}{}_k + {\frac 12}T_{mn}{}^i {\mathfrak H}^{mn}{}_k -  {\mathfrak E}_k{}^i &=& - {\mathfrak T}_k{}^i,\label{1st}\\
\check{\nabla}{}_l {\mathfrak H}^{kli}{}_j + {\frac 12}T_{mn}{}^k {\mathfrak H}^{mni}{}_j -  {\mathfrak E}^{ki}{}_j &=& {\mathfrak S}^i{}_j{}^k.\label{2nd}  
\end{eqnarray}
Here the generalized gravitational field momenta densities are introduced by
\begin{eqnarray}
{\mathfrak H}^{kli}{}_j := -\,2{\frac {\partial  {\mathfrak V}}{\partial R_{kli}{}^j}},\qquad
{\mathfrak H}^{ki}{}_j := -\,{\frac {\partial  {\mathfrak V}}{\partial T_{ki}{}^j}},\qquad
{\mathfrak M}^{kij} := -\,{\frac {\partial  {\mathfrak V}}{\partial Q_{kij}}},\label{HH}
\end{eqnarray}
and the gravitational hypermomentum density
\begin{equation}\label{EN}
{\mathfrak E}^{ki}{}_j := -  {\mathfrak H}^{ki}{}_j -  {\mathfrak M}^{ki}{}_j = -\,{\frac {\partial  {\mathfrak V}}{\partial N_{ki}{}^j}}.
\end{equation}
Furthermore, the generalized energy-momentum density of the gravitational field is
\begin{equation}
{\mathfrak E}_k{}^i = \delta_k^i  {\mathfrak V} + {\frac 12}Q_{kln}  {\mathfrak M}^{iln} + T_{kl}{}^n  {\mathfrak H}^{il}{}_n + R_{kln}{}^m  {\mathfrak H}^{iln}{}_m.\label{Eg}
\end{equation}
The sources of the gravitational field are the canonical energy-momentum tensor and the canonical hypermomentum densities of matter, respectively:
\begin{eqnarray}
{\mathfrak T}_k{}^i &:=& {\frac {\partial  {\mathfrak L}_{\rm mat}}{\partial\nabla_i\psi^A}}\,\nabla_k\psi^A - \delta^i_k {\mathfrak L}_{\rm mat}.\label{canD}\\
{\mathfrak S}^i{}_j{}^k &:=& {\frac {\partial  {\mathfrak L}_{\rm mat}}{\partial \Gamma_{ki}{}^j}} =  - {\frac {\partial  {\mathfrak L}_{\rm mat}}{\partial\nabla_k\psi^A}} \,(\sigma^A{}_B)_j{}^i \psi^B.\label{tD}
\end{eqnarray}

It is straightforward to verify that instead of the first field equation (\ref{1st}), one can use the so-called zeroth field equation which reads
\begin{equation}\label{0th}
2{\frac {\delta {\mathfrak V}}{\delta g_{ij}}} =  {\mathfrak t}^{ij}.
\end{equation}
On the right-hand side, the matter source is now represented by the metrical energy-momentum density which is defined by
\begin{equation}\label{tmet}
{\mathfrak t}_{ij} := 2{\frac {\partial {\mathfrak L}_{\rm mat}}{\partial g^{ij}}}.
\end{equation}
The system (\ref{1st}) and (\ref{2nd}) is completely equivalent to the system (\ref{2nd}) and (\ref{0th}), and it is a matter of convenience which one to solve. 

\subsection{From densities to tensors}

In actual metric-affine gravity models the Lagrangian densities are constructed in terms of the Lagrangian functions: ${\mathfrak L} = \sqrt{-g}L$, ${\mathfrak V} = \sqrt{-g}V$, ${\mathfrak L}_{\rm mat} = \sqrt{-g}L_{\rm mat}$. Accordingly, we find for the gravitational field momenta
\begin{equation}
{\mathfrak H}^{kli}{}_j = \sqrt{-g}H^{kli}{}_j,\qquad
{\mathfrak H}^{ki}{}_j = \sqrt{-g}H^{ki}{}_j,\qquad
{\mathfrak M}^{kij} = \sqrt{-g}M^{kij},\label{HHMt}
\end{equation}
and ${\mathfrak E}^{i}{}_j = \sqrt{-g}E^{i}{}_j$. The gravitational sources (\ref{canD}), (\ref{tD}) and (\ref{tmet}) are rewritten in terms of the canonical energy-momentum tensor, the canonical hypermomentum tensor, and the metrical energy-momentum tensor, respectively:
\begin{equation}
{\mathfrak T}_k{}^i = \sqrt{-g}\Sigma_k{}^i,\qquad {\mathfrak S}^i{}_j{}^k = \sqrt{-g}\Delta^i{}_j{}^k,\qquad {\mathfrak t}_{ij} = \sqrt{-g}t_{ij}.\label{TST}
\end{equation}
The usual spin arises as the antisymmetric part of the hypermomentum,
\begin{equation}
\tau_{ij}{}^k := \Delta_{[ij]}{}^k,\label{spindef}
\end{equation}
whereas the trace $\Delta^k = \Delta^i{}_i{}^k$ is the dilation current. The symmetric traceless part describes the proper hypermomentum \cite{Hehl:1995}.

As a result, the metric-affine field equations (\ref{1st}) and (\ref{2nd}) are recast into
\begin{eqnarray}
{\stackrel * \nabla}{}_nH^{in}{}_k + {\frac 12}T_{mn}{}^iH^{mn}{}_k - E_k{}^i &=& - \Sigma_k{}^i,\label{1st_t}\\
{\stackrel * \nabla}{}_lH^{kli}{}_j + {\frac 12}T_{mn}{}^kH^{mni}{}_j - E^{ki}{}_j &=& \Delta^i{}_j{}^k.\label{2nd_t}  
\end{eqnarray}

\subsubsection{Example: Matter model with microstructure}

In order to give an explicit example of physical matter with microstructure, we recall the hyperfluid model \cite{Obukhov:1993}. This is a direct generalization of the general relativistic ideal fluid variational theory \cite{Taub:1954,Schutz:1970} and of the spinning fluid model of Weyssenhoff and Raabe \cite{Weyssenhoff:1947,Obukhov:1987}. Using the variational principle for the hyperfluid \cite{Obukhov:1993}, one derives the canonical energy-momentum and hypermomentum tensors:
\begin{eqnarray}
\Sigma_k{}^i &=& \,v^iP_k - p\left(\delta_k^i - v^iv_k\right),\label{hypS}\\
\Delta^n{}_m{}^i &=& \,v^iJ_m{}^n,\label{hypD}
\end{eqnarray}
where $v^i$ is the 4-velocity of the fluid and $p$ is the pressure. Fluid elements are characterized by their microstructural properties: the momentum density $P_k$ and the intrinsic hypermomentum density $J_m{}^n$. 

\section{General Lagrange-Noether analysis}\label{Noether_sec}

As a first step, we notice that the gravitational (geometrical) and material variables can be described together by means of a multiplet, which we denote by $\Phi^J = (g_{ij},\Gamma_{ki}{}^j,\psi^A)$. We do not specify the range of the multi-index $J$ at this stage. The matter fields may include, besides the true material variables, also auxiliary fields such as Lagrange multipliers. With the help of the latter we can impose various constraints on the geometry of the spacetime. Furthermore, we can use the Lagrange multipliers to describe models in which the Lagrangian depends on arbitrary-order covariant derivatives of the curvature, torsion, and nonmetricity. Then the general action reads 
\begin{equation}
I = \int\,d^4x\,{\mathfrak L},\label{action}
\end{equation}
where the Lagrangian density ${\mathfrak L} = {\mathfrak L}(\Phi^J,\partial_i\Phi^J)$ depends arbitrarily on the set of fields $\Phi^J$ and their first derivatives. 

Our aim is to derive Noether identities that correspond to general coordinate transformations. However, it is more convenient to start with arbitrary infinitesimal transformations of the spacetime coordinates and the matter fields. They are given as follows:
\begin{eqnarray}
x^i &\longrightarrow& x'^i (x) = x^i + \delta x^i,\label{dx}\\
\Phi^J(x) &\longrightarrow& \Phi'^J(x') = \Phi^J(x) + \delta\Phi^J(x).
\label{dP}
\end{eqnarray} 
Within the present context it is not important whether this is a symmetry transformation under the action of any specific group. The total variation (\ref{dP}) is a result of the change of the form of the functions and of the change induced by the transformation of the spacetime coordinates (\ref{dx}). To distinguish the two pieces in the field transformation, it is convenient to introduce the {\it substantial variation}: 
\begin{equation}\label{sub}
\subsvar\Phi^J := \Phi'^J(x) - \Phi^J(x) = \delta\Phi^J - \delta x^k\partial_k\Phi^J.
\end{equation}
By definition, the substantial variation commutes with the partial derivative, $\subsvar\partial_i = \partial_i\subsvar$. 

We need the total variation of the action:
\begin{equation}
\delta I = \int \left[d^4x\,\delta {\mathfrak L} + \delta(d^4x)\,{\mathfrak L}\right].
\end{equation}
A standard derivation shows that under the action of the transformation (\ref{dx})-(\ref{dP}) the total variation reads
\begin{equation}
\delta I = \int d^4x \left[ {\frac {\delta {\mathfrak L}}{\delta\Phi^J}} \,\subsvar\Phi^J + \partial_i\left({\mathfrak L}\,\delta x^i + {\frac {\partial {\mathfrak L}}{\partial(\partial_i\Phi^J)}}\,\subsvar \Phi^J \right)\right]. \label{masterV}
\end{equation}
Here the variational derivative is defined, as usual, by
\begin{equation}\label{var}
{\frac {\delta {\mathfrak L}}{\delta\Phi^J}} := {\frac {\partial {\mathfrak L}}{\partial\Phi^J}} - \partial_i\left({\frac {\partial {\mathfrak L}}{\partial(\partial_i\Phi^J)}}\right).
\end{equation}

\subsection{General coordinate invariance}

Now we specialize to general coordinate transformations. For infinitesimal changes of the spacetime coordinates and (matter and gravity) fields (\ref{dx}) and (\ref{dP}) we have $x^i\rightarrow x^i + \delta x^i$, $g_{ij}\rightarrow g_{ij} + \delta g_{ij}$, $\Gamma_{ki}{}^j\rightarrow \Gamma_{ki}{}^j + \delta\Gamma_{ki}{}^j$, and $\psi^A \rightarrow \psi^A + \delta\psi^A$, where variations are given by (\ref{dex}), (\ref{dgij}), (\ref{dG}), and (\ref{dpsiA}). Substituting these variations into (\ref{masterV}), and making use of the substantial derivative definition (\ref{sub}), we find
\begin{eqnarray}
\delta I = -\,\int d^4x \biggl[\xi^k\,\Omega_k + (\partial_i\xi^k)\,\Omega_k{}^i
+ (\partial^2_{ij}\xi^k)\,\Omega_k{}^{ij} + (\partial^3_{ijn}\xi^k)\,\Omega_k{}^{ijn}\biggr],
\label{masterG} 
\end{eqnarray}
where explicitly
\begin{eqnarray}
\Omega_k &=& {\frac {\delta {\mathfrak L}}{\delta g_{ij}}}\,\partial_kg_{ij} + {\frac {\delta {\mathfrak L}}{\delta\psi^A}}\,\partial_k\psi^A + {\frac {\partial {\mathfrak L}}{\partial \Gamma_{ln}{}^m}}\partial_k\Gamma_{ln}{}^m + {\frac {\partial {\mathfrak L}}{\partial \partial_i\Gamma_{ln}{}^m}}\partial_k\partial_i\Gamma_{ln}{}^m\nonumber\\
&& + \,\partial_i\left({\frac {\partial {\mathfrak L}}{\partial \partial_ig_{mn}}}\partial_kg_{mn} + {\frac {\partial {\mathfrak L}}{\partial\partial_i\psi^A}} \,\partial_k\psi^A - \delta^i_k{\mathfrak L}\right),\label{Om1}\\
\Omega_k{}^i &=& 2{\frac {\delta {\mathfrak L}}{\delta g_{ij}}}\,g_{kj} + {\frac {\delta {\mathfrak L}}{\delta\psi^A}}\,(\sigma^A{}_B)_k{}^i\,\psi^B + {\frac {\partial {\mathfrak L}}{\partial\partial_i\psi^A}}\partial_k\psi^A  - \delta^i_k{\mathfrak L}\nonumber\\ 
&& + {\frac {\partial {\mathfrak L}}{\partial \partial_ig_{mn}}}\partial_kg_{mn} + \partial_j\left(2{\frac {\partial {\mathfrak L}}{\partial \partial_jg_{in}}}g_{nk} + {\frac {\partial {\mathfrak L}}{\partial\partial_j\psi^A}}(\sigma^A{}_B)_k{}^i\psi^B\right)  \nonumber\\
&& + \,{\frac {\partial {\mathfrak L}}{\partial \Gamma_{li}{}^j}}\,\Gamma_{lk}{}^j + {\frac {\partial {\mathfrak L}}{\partial \Gamma_{il}{}^j}}\,\Gamma_{kl}{}^j - {\frac {\partial {\mathfrak L}}{\partial \Gamma_{lj}{}^k}}\,\Gamma_{lj}{}^i + \,{\frac {\partial {\mathfrak L}}{\partial \partial_i\Gamma_{ln}{}^m}}\,\partial_k \Gamma_{ln}{}^m \nonumber\\
&&  + {\frac {\partial {\mathfrak L}}{\partial \partial_n\Gamma_{il}{}^m}} \,\partial_n\Gamma_{kl}{}^m + {\frac {\partial {\mathfrak L}}{\partial \partial_n\Gamma_{li}{}^m}}\,\partial_n \Gamma_{lk}{}^m - {\frac {\partial {\mathfrak L}}{\partial \partial_n\Gamma_{lm}{}^k}}\,\partial_n\Gamma_{lm}{}^i ,\label{Om2}\\
\Omega_k{}^{ij} &=& {\frac {\partial {\mathfrak L}}{\partial\partial_{(i}\psi^A}} (\sigma^A{}_B)_k{}^{j)}\psi^B + {\frac {\partial {\mathfrak L}}{\partial \Gamma_{(ij)}{}^k}}
+ {\frac {\partial {\mathfrak L}}{\partial \partial_{(i}\Gamma_{j)l}{}^m}}\Gamma_{kl}{}^m \nonumber\\
&&+ \,2{\frac {\partial {\mathfrak L}}{\partial \partial_{(i}g_{j)n}}}g_{kn} + {\frac {\partial {\mathfrak L}}{\partial \partial_{(i}\Gamma_{|l|j)}{}^m}}\,\Gamma_{lk}{}^m - {\frac {\partial {\mathfrak L}}{\partial \partial_{(i}\Gamma_{|ln|}{}^k}}\,\Gamma_{ln}{}^{j)}, \label{Om3}\\
\Omega_k{}^{ijn} &=& {\frac {\partial {\mathfrak L}}{\partial \partial_{(n}\Gamma_{ij)}{}^k}}.\label{Om4}
\end{eqnarray}
If the action is invariant under general coordinate transformations, $\delta I = 0$, in view of the arbitrariness of the function $\xi^i$ and its derivatives, we find the set of four Noether identities:
\begin{equation}\label{NoeG}
\Omega_k = 0,\quad \Omega_k{}^i = 0,\quad \Omega_k{}^{ij} = 0, \quad \Omega_k{}^{ijn} = 0.
\end{equation}

General coordinate invariance is a natural consequence of the fact that the action (\ref{action}) and the Lagrangian ${\mathfrak L}$ are constructed only from covariant objects. Namely, ${\mathfrak L} = {\mathfrak L}(\psi^A, \nabla_i\psi^A, g_{ij}, R_{kli}{}^j, N_{kj}{}^i)$ is a function of the metric, the curvature (\ref{curv}), the torsion (\ref{tors}), the matter fields, and their covariant derivatives (\ref{Dpsi}). Denoting 
\begin{eqnarray}
\rho^{ijk}{}_l := {\frac {\partial {\mathfrak L}}{\partial R_{ijk}{}^l}},\qquad
\mu^{ij}{}_k := {\frac {\partial {\mathfrak L}}{\partial N_{ij}{}^k}},\label{dLRN}
\end{eqnarray}
we find for the derivatives of the Lagrangian
\begin{eqnarray}
{\frac {\partial {\mathfrak L}}{\partial \Gamma_{ij}{}^k}} &=& -\,{\frac {\partial {\mathfrak L}} {\partial \nabla_i\psi^A}}(\sigma^A{}_B)_k{}^j\,\psi^B - \mu^{ij}{}_k\nonumber\\
&& + 2\rho^{inl}{}_k\Gamma_{nl}{}^j + 2\rho^{nij}{}_l\Gamma_{nk}{}^l,\label{dLG}\\
{\frac {\partial {\mathfrak L}}{\partial \partial_i\Gamma_{jk}{}^l}} &=& 2\rho^{ijk}{}_l,\label{dLdG}\\ 
{\frac {\partial {\mathfrak L}}{\partial \partial_kg_{ij}}} &=& {\frac 12}\left(\mu^{(ki)j} + \mu^{(kj)i} - \mu^{(ij)k}\right).\label{dLdm}
\end{eqnarray}
As a result, we straightforwardly verify that $\Omega_k{}^{ij} = 0$ and $\Omega_k{}^{ijn} = 0$ are indeed satisfied identically. 

Using (\ref{dLG}) and (\ref{dLdG}), we then recast the two remaining Noether identities (\ref{Om1}) and (\ref{Om2}) into
\begin{eqnarray}
\hspace{-0.2cm}\Omega_k &=& {\frac {\delta {\mathfrak L}}{\delta g_{ij}}}\,\partial_kg_{ij} + {\frac {\delta {\mathfrak L}}{\delta\psi^A}}\,\partial_k\psi^A + \partial_i\!\left(\!{\frac {\partial {\mathfrak L}}{\partial\nabla_i\psi^A}}\nabla_k\psi^A - \delta^i_k{\mathfrak L} \!\right)\!\nonumber\\
&& +\,\widehat{\nabla}{}_j\!\left(\!{\frac {\partial {\mathfrak L}}{\partial\nabla_j\psi^A}} \,(\sigma^A{}_B)_m{}^n\,\psi^B\!\right)\!\Gamma_{kn}{}^m + \rho^{iln}{}_m\partial_kR_{iln}{}^m\nonumber\\
&& + \,{\frac {\partial {\mathfrak L}}{\partial\nabla_l\psi^A}}\,(\sigma^A{}_B)_m{}^n \,\psi^B\,R_{lkn}{}^m + \mu^{ln}{}_m\partial_kN_{ln}{}^m \nonumber\\
&& +\,{\frac 12}\check{\nabla}_i\left(\mu^{(im)n} + \mu^{(in)m} - \mu^{(mn)i}\right)\partial_k g_{mn},\label{Om1a}\\
\hspace{-0.2cm}\Omega_k{}^i &=& 2{\frac {\delta {\mathfrak L}}{\delta g_{ij}}}\,g_{kj} + {\frac {\delta {\mathfrak L}}{\delta\psi^A}}\,(\sigma^A{}_B)_k{}^i\,\psi^B + {\frac {\partial {\mathfrak L}}{\partial\nabla_i\psi^A}}\nabla_k\psi^A - \delta^i_k{\mathfrak L}  \nonumber\\ 
&& + \widehat{\nabla}{}_j\left({\frac {\partial {\mathfrak L}}{\partial\nabla_j\psi^A}}(\sigma^A{}_B)_k{}^i\psi^B\right) - \mu^{ln}{}_kN_{ln}{}^i + \mu^{il}{}_nN_{kl}{}^n + \mu^{li}{}_nN_{lk}{}^n \nonumber\\
&& + \,2\rho^{iln}{}_mR_{kln}{}^m + \rho^{lni}{}_mR_{lnk}{}^m - \rho^{lnm}{}_kR_{lnm}{}^i\nonumber\\
&& +\,\check{\nabla}_n\left(\mu^{(ni)j} + \mu^{(nj)i} - \mu^{(ij)n}\right)g_{jk}= 0.\label{Om2a}
\end{eqnarray}

Notice that the variational derivative (\ref{var}) with respect to the matter fields can be identically rewritten as
\begin{equation}\label{covar}
{\frac {\delta {\mathfrak L}}{\delta\psi^A}} \equiv {\frac {\partial {\mathfrak L}}{\partial\psi^A}} - \widehat{\nabla}{}_j\left({\frac {\partial {\mathfrak L}}{\partial\nabla_j\psi^A}}\right),
\end{equation}
and turns out to be a covariant tensor density. It is also worthwhile to note, that the variational derivative w.r.t.\ the metric is explicitly a covariant density. This follows from the fact that the Lagrangian depends on $g_{ij}$ not only directly, but also through the objects $Q_{kij}$ and $N_{ki}{}^j$. Taking this into account, we find
\begin{eqnarray}
{\frac {\delta {\mathfrak L}}{\delta g_{ij}}} &=& {\frac {d {\mathfrak L}}{d g_{ij}}} - \partial_n\left(
{\frac {\partial {\mathfrak L}}{\partial \partial_ng_{ij}}}\right) \nonumber\\
&=& {\frac {\partial {\mathfrak L}}{\partial g_{ij}}}- {\frac 12}\check{\nabla}_n\left(\mu^{(ni)j} + \mu^{(nj)i} - \mu^{(ij)n}\right).\label{dLgij}
\end{eqnarray}

The Noether identity (\ref{Om2a}) is a covariant relation. In contrast, (\ref{Om1a}) is apparently non-covariant. However, this can be easily repaired by replacing $\Omega_k = 0$ with an equivalent covariant Noether identity: $\Omega'{}_k := \Omega_k -  \Gamma_{kn}{}^m\Omega_m{}^n = 0$. Explicitly, we find
\begin{eqnarray}
\Omega'{}_k &=& {\frac {\delta {\mathfrak L}}{\delta\psi^A}}\,\nabla_k\psi^A + \widehat{\nabla}{}_i\!\left(\!{\frac {\partial {\mathfrak L}}{\partial\nabla_i\psi^A}} \,\nabla_k\psi^A - \delta^i_k{\mathfrak L} \!\right)\nonumber\\ 
&&  - \left({\frac {\partial {\mathfrak L}}{\partial\nabla_i\psi^A}} \,\nabla_l\psi^A - \delta^i_l{\mathfrak L}\right) T_{ki}{}^l + {\frac {\partial {\mathfrak L}}{\partial\nabla_l\psi^A}}\,(\sigma^A{}_B)_m{}^n \,\psi^B\,R_{lkn}{}^m \nonumber\\
&& + \left[ - {\frac {\delta {\mathfrak L}}{\delta g_{ij}}} - {\frac 12}\check{\nabla}_n\left(\mu^{(ni)j} + \mu^{(nj)i} - \mu^{(ij)n}\right)\right]Q_{kij} \nonumber\\
&& + \,\mu^{ln}{}_m\nabla_kN_{ln}{}^m + \rho^{iln}{}_m\nabla_kR_{iln}{}^m   \nonumber\\ 
&= & 0.\label{Om1b}
\end{eqnarray}

On-shell, i.e.\ assuming that the matter fields satisfy the field equations ${\delta {\mathfrak L}}/{\delta\psi^A} = 0$, the Noether identities (\ref{Om2a}) and (\ref{Om1b}) reduce to the {\it conservation laws} for the energy-momentum and hypermomentum, respectively.

Equation (\ref{Om2a}) contains a relation between the canonical and the metrical energy-momentum tensor, and the conservation law of the hypermomentum. In the next section we turn to the discussion of models with general nonminimal coupling.

\section{Conservation laws in models with nonminimal coupling}\label{conservation_sec}

The results obtained in the previous section are applicable to {\it any} theory in which the Lagrangian depends arbitrarily on the matter fields and the gravitational field strengths. Now we specialize to the class of models described by an interaction Lagrangian of the form
\begin{equation}
L = F(g_{ij},R_{kli}{}^j, N_{kl}{}^i)\,L_{\rm mat}(\psi^A, \nabla_i\psi^A).\label{LFL}
\end{equation}
Here $L_{\rm mat}(\psi^A, \nabla_i\psi^A)$ is the ordinary matter Lagrangian. We call  $F = F(g_{ij},R_{kli}{}^j, N_{kl}{}^i)$ the coupling function and assume that it can depend arbitrarily on its arguments, i.e., on all covariant gravitational field variables of MAG. When $F = 1$ we recover the minimal coupling case. 

\subsection{Identities for the nonminimal coupling function}\label{Fidentities_subsec}

As a preliminary step, let us derive identities which are satisfied for the nonminimal coupling function $F = F(g_{ij},R_{kli}{}^j, N_{kl}{}^i)$. For this, we apply the above Lagrange-Noether machinery to the auxiliary Lagrangian density ${\mathfrak L}_0 = \sqrt{-g}\,F$. This quantity does not depend on the matter fields, and both (\ref{Om2a}) and (\ref{Om1b}) are considerably simplified. In particular, we have
\begin{equation}
{\frac {\delta {\mathfrak L}_0}{\delta g_{ij}}} = \sqrt{-g}\left({\frac 12}Fg^{ij} + F^{ij} \right),\qquad F^{ij} := {\frac {\delta F}{\delta g_{ij}}}.\label{dFg}
\end{equation}
Then we immediately see that (\ref{Om2a}) and (\ref{Om1b}) reduce to
\begin{eqnarray}
\nabla_k F &\equiv& \left[-F^{ij} - {\frac 12}\widetilde{\nabla}_n\left({\stackrel 0 \mu}{}^{(ni)j} + {\stackrel 0 \mu}{}^{(nj)i} - {\stackrel 0 \mu}{}^{(ij)n}\right)\right]Q_{kij} \nonumber\\
&& + \,{\stackrel 0 \rho}{}^{iln}{}_m\nabla_kR_{iln}{}^m + {\stackrel 0 \mu}{}^{ln}{}_m\nabla_kN_{ln}{}^m,\label{F1}\\
2F_k{}^i &\equiv& - \,2{\stackrel 0 \rho}{}^{iln}{}_mR_{kln}{}^m - {\stackrel 0 \rho}{}^{lni}{}_m R_{lnk}{}^m + {\stackrel 0 \rho}{}^{lnm}{}_kR_{lnm}{}^i\nonumber\\
&& - \,{\stackrel 0 \mu}{}^{ln}{}_kN_{ln}{}^i  + {\stackrel 0 \mu}{}^{il}{}_nN_{kl}{}^n + {\stackrel 0 \mu}{}^{li}{}_nN_{lk}{}^n\nonumber\\
&& +\,\widetilde{\nabla}_n\left({\stackrel 0 \mu}{}^{(ni)j} + {\stackrel 0 \mu}{}^{(nj)i} - {\stackrel 0 \mu}{}^{(ij)n}\right)g_{jk}.\label{F2}
\end{eqnarray}
Here we denoted
\begin{equation}\label{dFRT}
{\stackrel 0 \rho}{}^{ijk}{}_l := {\frac {\partial F}{\partial R_{ijk}{}^l}},\qquad
{\stackrel 0 \mu}{}^{ij}{}_k := {\frac {\partial F}{\partial N_{ij}{}^k}}.
\end{equation}
Notice that for any tensor density we have relations (\ref{nablatilde}) and (\ref{nablastar}).

The identity (\ref{F1}) is naturally interpreted as a generally covariant generalization of the chain rule for the total derivative of a function of several variables. This becomes obvious when we notice that (\ref{dLgij}) implies
\begin{eqnarray}
\left[-F^{ij} - {\frac 12}\widetilde{\nabla}_n\left({\stackrel 0 \mu}{}^{(ni)j} + {\stackrel 0 \mu}{}^{(nj)i} - {\stackrel 0 \mu}{}^{(ij)n}\right)\right]Q_{kij} 
= {\frac {\partial F} {\partial g_{ij}}}\,\nabla_kg_{ij}.\label{difFg}
\end{eqnarray}
It should be stressed that (\ref{F1}) and (\ref{F2}) are true identities, they are satisfied for any function $F(g_{ij}, R_{kli}{}^j, N_{kl}{}^i)$ irrespectively of the field equations that can be derived from the corresponding action. 

\subsection{Conservation laws}\label{Conslaws_subsec}

Now we are in a position to derive the conservation laws for the general nonminimal coupling model (\ref{LFL}). Recall the definitions of the canonical energy-momentum tensor (\ref{canD}), the canonical hypermomentum tensor (\ref{tD}) and the metrical energy-momentum tensor (\ref{tmet}). 

In view of the product structure of the Lagrangian (\ref{LFL}), the derivatives are easily evaluated, and the conservation laws (\ref{Om2a}) and (\ref{Om1b}) reduce to
\begin{eqnarray}
&& -\,Ft_k{}^i - {\stackrel * \nabla}{}_n\left(F\Delta^i{}_k{}^n\right) + F\Sigma_k{}^i + \Bigl[2F_k{}^i \nonumber\\
&& +\,2{\stackrel 0 \rho}{}^{iln}{}_mR_{kln}{}^m + {\stackrel 0 \rho}{}^{lni}{}_m R_{lnk}{}^m - {\stackrel 0 \rho}{}^{lnm}{}_kR_{lnm}{}^i \nonumber\\ 
&& +\,{\stackrel 0 \mu}{}^{ln}{}_kN_{ln}{}^i  - {\stackrel 0 \mu}{}^{il}{}_nN_{kl}{}^n - {\stackrel 0 \mu}{}^{li}{}_nN_{lk}{}^n  \nonumber\\ 
&& -\,\widetilde{\nabla}_n\left({\stackrel 0 \mu}{}^{(ni)j} + {\stackrel 0 \mu}{}^{(nj)i} - {\stackrel 0 \mu}{}^{(ij)n}\right)g_{jk}\Bigr]L_{\rm mat} = 0, \label{cons1a}\\
&& {\stackrel * \nabla}{}_i\left(F\Sigma_k{}^i\right)  + F\Bigl[ - \Sigma_l{}^i T_{ki}{}^l + \Delta^m{}_n{}^l R_{klm}{}^n +  {\frac 12}t^{ij}Q_{kij}\Bigr] \nonumber\\
&& + \Bigl\{{\stackrel 0 \rho}{}^{iln}{}_m\nabla_kR_{iln}{}^m + {\stackrel 0 \mu}{}^{ln}{}_m\nabla_kN_{ln}{}^m + \Bigl[-F^{ij} \nonumber\\
&&- {\frac 12}\widetilde{\nabla}_n\!\Bigl({\stackrel 0 \mu}{}^{(ni)j} + {\stackrel 0 \mu}{}^{(nj)i} - {\stackrel 0 \mu}{}^{(ij)n}\Bigr)\Bigr]Q_{kij}\Bigr\}L_{\rm mat} = 0.\label{cons2a}
\end{eqnarray}
After we take into account the identities (\ref{F1}) and (\ref{F2}), the conservation laws (\ref{cons1a}) and (\ref{cons2a}) are brought to the final form:
\begin{eqnarray}
F\Sigma_k{}^i &=& Ft_k{}^i + {\stackrel * \nabla}{}_n\left(F\Delta^i{}_k{}^n\right),\label{cons1b}\\
{\stackrel * \nabla}{}_i\left(F\Sigma_k{}^i\right) &=& F \left( \Sigma_l{}^i T_{ki}{}^l - \Delta^m{}_n{}^l R_{klm}{}^n  - {\frac 12}t^{ij}Q_{kij} \right) - L_{\rm mat}\nabla_kF.\label{cons2b}
\end{eqnarray}
Lowering the index in (\ref{cons1b}) and antisymmetrizing, we derive the conservation law for the spin
\begin{equation}\label{skewS}
F\Sigma_{[ij]} + {\stackrel * \nabla}{}_n\left(F\tau_{ij}{}^n\right) + Q_{nl[i}\Delta^l{}_{j]}{}^n = 0.
\end{equation}
This is a generalization of the usual conservation law of the total angular momentum for the case of nonminimal coupling. 

\subsection{Rewriting the conservation laws}\label{nonmin_sec}

Using the definition (\ref{dstar}) and decomposing the connection into the Riemannian and non-Riemannian parts (\ref{dist}), we can recast the conservation law (\ref{cons1b}) into an equivalent form
\begin{equation}
\widetilde{\nabla}_j(F\Delta^i{}_k{}^j) =  F(\Sigma_k{}^i - t_k{}^i + N_{nm}{}^i\Delta^m{}_k{}^n - N_{nk}{}^m\Delta^i{}_m{}^n).\label{cons1c}
\end{equation}
In a similar way we can rewrite the conservation law (\ref{cons2b}). At first, with the help of (\ref{TN}) and (\ref{QN}) we notice that
\begin{equation}
F \left( \Sigma_l{}^i T_{ki}{}^l - {\frac 12}t^{ij}Q_{kij} \right) = F(t_l{}^i - \Sigma_l{}^i)N_{ki}{}^l + F\Sigma_l{}^i N_{ik}{}^l.\label{TTN}
\end{equation}
Then substituting here (\ref{cons1c}) and making use of (\ref{dist}), (\ref{dstar}), and the curvature decomposition (\ref{RRN}), after some algebra we recast (\ref{cons2b}) into
\begin{equation}\label{cons2c}
\widetilde{\nabla}_j\{F(\Sigma_k{}^j + \Delta^m{}_n{}^j N_{km}{}^n)\} = -\,F\Delta^m{}_n{}^i(\widetilde{R}_{kim}{}^n - \widetilde{\nabla}_kN_{im}{}^n) - L_{\rm mat}\widetilde{\nabla}_kF.
\end{equation}
For the minimal coupling case, such a conservation law was derived in \cite{Neeman:1997,Obukhov:2006}. The importance of this form of the energy-momentum conservation law lies in the clear separation of the Riemannian and non-Riemannian geometrical variables. As we see, the post-Riemannian geometry enters (\ref{cons2c}) only in the form of the distortion tensor $N_{kj}{}^i$ which is coupled only to the hypermomentum current $\Delta^m{}_n{}^i$. This means that, in the {\it minimal coupling} case, ordinary matter -- i.e.\ without microstructure, $\Delta^m{}_n{}^i=0$ -- does {\it not} couple to the non-Riemannian geometry. In contrast, in the {\it nonminimal coupling} case, the derivative of the coupling coupling function $F$ on the right-hand side of (\ref{cons2c}) may lead to a coupling between non-Riemannian structures and ordinary matter.    

\subsection{Conserved current induced by a spacetime symmetry}\label{Induced_subsec}

Every generalized Killing vector field $\zeta^k$ generates a conserved current. This can be demonstrated from the analysis of the system (\ref{cons1c}) and (\ref{cons2c}) as follows. Let us contract equation (\ref{cons1c}) with $\widetilde{\nabla}_i\zeta^k$ and equation (\ref{cons2c}) with $\zeta^k$, then subtract the resulting expressions. Note that the contraction $t_k{}^i\widetilde{\nabla}_i\zeta^k = 0$ vanishes because the first factor is a symmetric tensor and the second one is skew-symmetric. Then after some algebra we derive
\begin{equation}
\widetilde{\nabla}_i\,{\stackrel \zeta I}{}^i = F\Delta^m{}_n{}^i{\cal L}_\zeta N_{im}{}^n - L_{\rm mat}{\cal L}_\zeta F.\label{DI1}
\end{equation}
Here we associate a current with a Killing vector field via
\begin{equation}
{\stackrel \zeta I}{}^i := F\left[\zeta^k\Sigma_k{}^i - (\overline{\nabla}_m\zeta^n)\Delta^m{}_n{}^i\right].\label{I1}
\end{equation}
Note that the transposed connection appears here. 
The right-hand side of (\ref{DI1}) depends linearly on the Lie derivatives along the Killing vector: ${\cal L}_\zeta F = \zeta^k\widetilde{\nabla}_kF$ and 
\begin{equation}\label{LzN}
{\cal L}_\zeta N_{im}{}^n = \zeta^k\widetilde{\nabla}_kN_{im}{}^n + (\widetilde{\nabla}_i\zeta^k)N_{km}{}^n + (\widetilde{\nabla}_m\zeta^k)N_{ik}{}^n - (\widetilde{\nabla}_k\zeta^n)N_{im}{}^k.
\end{equation}

When $\zeta^k$ is a {\it generalized Killing} vector, we have ${\cal L}_\zeta N_{im}{}^n = 0$ in view of (\ref{LieN}). Furthermore, recalling that $F = F(g_{ij}, R_{klj}{}^i, N_{kj}{}^i)$, we find 
\begin{equation}
{\cal L}_\zeta F = {\frac {\partial F}{\partial g_{ij}}}{\cal L}_\zeta g_{ij} + {\frac {\partial F}{\partial R_{klj}{}^i}}{\cal L}_\zeta R_{klj}{}^i + {\frac {\partial F}{\partial N_{kj}{}^i}}{\cal L}_\zeta N_{kj}{}^i = 0,\label{LieF}
\end{equation}
by making use of (\ref{LieN}), (\ref{LgMAG}), and (\ref{LRMAG}). 

As a result, the right-hand side of (\ref{DI1}) vanishes for the generalized Killing vector field, and we conclude that the induced current (\ref{I1}) is conserved
\begin{equation}
\widetilde{\nabla}_i\,{\stackrel \zeta I}{}^i = 0.\label{DI2}
\end{equation}
This generalizes the earlier results reported in \cite{Trautman:1973,Obukhov:Rubilar:2006,Obukhov:Rubilar:2007}.
In section \ref{cons_mom} we will show that there is a conserved quantity constructed from the multipole moments which is a direct counterpart of the induced current (\ref{I1}). It is worthwhile to give an equivalent form of the latter:
\begin{equation}
{\stackrel \zeta I}{}^i = F\left[\zeta^k(\Sigma_k{}^i + \Delta^m{}_n{}^iN_{km}{}^n) - (\widetilde{\nabla}_m\zeta^n)\Delta^m{}_n{}^i\right].\label{I2}
\end{equation}

\subsection{Riemannian limit}\label{Riemannian_subsec}

Our results contain the Riemannian theory as a special case. Suppose that the torsion and the nonmetricity are absent $T_{ij}{}^k = 0$, $Q_{kij} = 0$, hence $N_{ij}{}^k = 0$. Then for usual matter without microstructure (i.e.\ matter with $\Delta^m{}_n{}^i = 0$) the canonical and the metrical energy-momentum tensors coincide, $\Sigma_k{}^i = t_k{}^i$. As a result, the conservation law (\ref{cons2b}) reduces to
\begin{equation}
\widetilde{\nabla}_i t_k{}^i = {\frac 1F}\left(- L_{\rm mat}\delta_k^i -t_k{}^i\right)\widetilde{\nabla}_iF.\label{consF}
\end{equation}
This conservation law for the general nonminimal coupling model was derived earlier in \cite{Puetzfeld:Obukhov:2013} without using the Noether theorem, directly from the field equations\footnote{Notice a different conventional sign, as compared to our previous work \cite{Puetzfeld:Obukhov:2013}.}. The old result established the conservation law for the case in which $F = F(\widetilde{R}_{ijk}{}^l)$ depends arbitrarily on the components of the curvature tensor, correcting some erroneous derivations in the literature, see \cite{Puetzfeld:Obukhov:2013} for details. 

Quite remarkably, (\ref{consF}) generalizes the earlier result to the case in which the nonminimal coupling function $F$ is a general scalar function of the curvature tensor. 

\section{General multipolar framework}\label{master_sec}

In this section we derive ``master equations of motion'' for a general extended test body, which is characterized by a set of currents 
\begin{equation}
J^{Aj}.\label{JA}
\end{equation}
Normally, these are so-called Noether currents that correspond to an invariance of the action under a certain symmetry group. However, this is not necessary, and any set of currents is formally allowed. We call $J^{Aj}$ dynamical currents. The generalized index (capital Latin letters $A,B,\dots$) labels different components of the currents. 

As the starting point for the derivation of the equations of motion for generalized multipole moments, we consider the following conservation law:
\begin{equation}
\widetilde{\nabla}_jJ^{Aj} = -\,\Lambda_{jB}{}^A\,J^{Bj} - \Pi^A{}_{\dot{B}}K^{\dot{B}}.\label{dJ}
\end{equation}
On the right-hand side, we introduce objects that can be called material currents 
\begin{equation}
K^{\dot{A}}\label{Xi}
\end{equation}
to distinguish them from the dynamical currents $J^{Aj}$. The number of components of the dynamical and material currents is different, hence we use a different index with dot $\dot{A}, \dot{B}, \dots$,  the range of which does not coincide with that of $A,B,\dots$. At this stage we do not specify the ranges of both types of indices, this will be done for the particular examples which we analyze later. As usual, Einstein's summation rule over repeated indices is assumed for the generalized indices as well as for coordinate indices. 

Both sets of currents $J^{Aj}$ and $K^{\dot{A}}$ are constructed from the variables that describe the structure and the properties of matter inside the body. In contrast, the objects 
\begin{equation}
\Lambda_{jB}{}^A,\qquad \Pi^A{}_{\dot{B}},
\end{equation}
do not depend on the matter, but they are functions of the external classical fields which act on the body and thereby determine its motion. The list of such external fields include the electromagnetic, gravitational and scalar fields.

We will now derive the equations of motion of a test body by utilizing the covariant expansion method of Synge \cite{Synge:1960}. For this we need the following auxiliary formula for the absolute derivative of the integral of an arbitrary bitensor density ${\mathfrak B}^{x_1 y_1}={\mathfrak B}^{x_1 y_1}(x,y)$ (the latter is a tensorial function of two spacetime points):
\begin{eqnarray}
{\frac{D}{ds}} \int\limits_{\Sigma(s)}{\mathfrak B}^{x_1 y_1} d \Sigma_{x_1} = \int\limits_{\Sigma(s)} \widetilde{\nabla}_{x_1}{\mathfrak B}^{x_1 y_1} w^{x_2} d \Sigma_{x_2} + \int\limits_{\Sigma(s)} v^{y_2} \widetilde{\nabla}_{y_2} {\mathfrak B}^{x_1 y_1} d \Sigma_{x_1}.\label{int_aux}
\end{eqnarray}
Here $v^{y_1}:=dx^{y_1}/ds$, $s$ is the proper time, ${\frac{D}{ds}} = v^i\widetilde{\nabla}_i$, and the integral is performed over a spatial hypersurface. Note that in our notation the point to which the index of a bitensor belongs can be directly read from the index itself; e.g., $y_{n}$ denotes indices at the point $y$. Furthermore, we will now associate the point $y$ with the world-line of the test body under consideration. 
Here $\sigma$ denotes Synge's \cite{Synge:1960} world-function and $\sigma^y$ its first covariant derivative; $g^y{}_x$ is the parallel propagator for vectors. For objects with more complicated tensorial properties the parallel propagator is straightforwardly generalized to $G^Y{}_X$ and $G^{\dot{Y}}{}_{\dot{X}}$. We will need these generalized propagators to deal with the dynamical and material currents $J^{Aj}$ and $K^{\dot{A}}$. More details are collected in the appendix on our conventions. 

After these preliminaries, we introduce integrated moments for the two types of currents via (for $n = 0,1,\dots)$
\begin{eqnarray}
j^{y_1\cdots y_n Y_0}  \!&=&\! (-1)^n\!\!\!\!\int\limits_{\Sigma(\tau)}\!\!\!\sigma^{y_1}\!\cdots\!\sigma^{y_n}G^{Y_0}{}_{X_0}{\mathfrak J}^{X_0 x''}d\Sigma_{x''},\label{j1n}\\
i^{y_1\dots y_{n} Y_0 y'} \!&=&\! (-1)^n\!\!\!\!\int\limits_{\Sigma(\tau)}\!\!\!\sigma^{y_1}\!\cdots\!\sigma^{y_n}G^{Y_0}{}_{X_0}g^{y'}{}_{x'}{\mathfrak J}^{X_0 x'}w^{x''}d\Sigma_{x''}, \label{i1n}\\
m^{y_1\dots y_{n} \dot{Y}_0 } \!&=&\! (-1)^n\!\!\!\!\int\limits_{\Sigma(\tau)}\!\!\!\sigma^{y_1}\!\cdots\!\sigma^{y_n}G^{\dot{Y}_0}{}_{\dot{X}_0}{\mathfrak K}^{\dot{X}_0}w^{x''}d\Sigma_{x''}.\label{m1n}
\end{eqnarray}
Here the densities are constructed from the currents: ${\mathfrak J}^{X_0 x_1} = \sqrt{-g}J^{X_0 x_1}$ and ${\mathfrak K}^{\dot{X}_0} = \sqrt{-g}K^{\dot{X}_0}$. 
Integrating (\ref{dJ}) and making use of (\ref{int_aux}), we find the following ``master equation of motion'' for the generalized multipole moments:
\begin{eqnarray}
{\frac{D}{ds}} j^{y_1\cdots y_n Y_0}  \!&=&\! - n\, v^{(y_1} j^{y_2 \dots y_n) Y_0} + n\, i^{(y_1 \dots y_{n-1}|Y_0|y_n)} \nonumber \\
&&- \gamma^{Y_0}{}_{Y'y''y_{n+1}}\left(i^{y_1 \dots y_{n}Y'y''} + j^{y_1 \dots y_{n}Y'}v^{y''}\right)\nonumber\\
&&- \Lambda_{y'Y''}{}^{Y_0}i^{y_1 \dots y_{n}Y''y'} - \Lambda_{y'Y''}{}^{Y_0}{}_{;y_{n+1}}i^{y_1 \dots y_{n+1}Y''y'} \nonumber \\
&&- \Pi^{Y_0}{}_{\dot{Y}'}m^{y_1 \dots y_{n}\dot{Y}'} - \Pi^{Y_0}{}_{\dot{Y}';y_{n+1}}m^{y_1 \dots y_{n+1}\dot{Y}'}\nonumber\\ 
&&+ \sum\limits^{\infty}_{k=2}{\frac 1{k!}}\Bigl[-(-1)^k n\, \alpha^{(y_1}{}_{y' y_{n+1} \dots y_{n+k}} i^{y_2 \dots y_n)y_{n+1} \dots y_{n+k} Y_0 y'} \nonumber \\
&& + (-1)^k n\, v^{y'} \beta^{(y_1}{}_{y' y_{n+1} \dots y_{n+k}} j^{y_2 \dots y_n)y_{n+1} \dots y_{n+k} Y_0}\nonumber\\ 
&&+ (-1)^k\gamma^{Y_0}{}_{Y'y''y_{n+1}\dots y_{n+k}}\left(i^{y_1 \dots y_{n+k}Y'y''} + j^{y_1 \dots y_{n+k}Y'}v^{y''}\right) \nonumber \\
&&- \Lambda_{y'Y''}{}^{Y_0}{}_{;y_{n+1}\dots y_{n+k}}i^{y_1 \dots y_{n+k}Y''y'}\nonumber\\ 
&&- \Pi^{Y_0}{}_{\dot{Y}';y_{n+1}\dots y_{n+k}}m^{y_1 \dots y_{n+k}\dot{Y}'}\Bigr].\label{master}
\end{eqnarray}

\subsection{Electrodynamics in Minkowski spacetime}\label{eom_max}

To see how the general formalism works, let us consider the motion of electrically charged extended test bodies under the influence of the electromagnetic field in flat Minkowski spacetime. This problem was earlier analyzed by means of a different approach in \cite{Dixon:1967}. 

In this case, it is convenient to recast the set of dynamical currents into the form of a column
\begin{equation}
J^{Aj} = \left(\begin{array}{c}J^j \\ \Sigma^{kj}\end{array}\right),\label{Jmax}
\end{equation}
where $J^j$ is the electric current and $\Sigma^{kj}$ is the energy-momentum tensor. Physically, the structure of the dynamical current is crystal clear: the matter elements of an extended body are characterized by the two types of ``charges'', the electrical charge (the upper component) and the mass (the lower component). 

The generalized conservation law comprises two components of different tensor dimension:
\begin{equation}
\widetilde{\nabla}_j \left(\begin{array}{c}J^j \\ \Sigma^{kj}\end{array}\right) = 
\left(\begin{array}{c}0 \\ - F^{kj}J_j\end{array}\right),\label{dJmax}
\end{equation}
where the lower component of the right-hand side describes the usual Lorentz force. 

Accordingly, we indeed recover for the dynamical current (\ref{Jmax}) the conservation law in the form (\ref{dJ}), where $K^{\dot{B}} = 0$ and
\begin{equation}
\Lambda_{jB}{}^A = \left(\begin{array}{c|c}0 & 0\\ \hline F_j{}^k & 0\end{array}\right).\label{LamMax}
\end{equation}
The generalized moments (\ref{j1n})-(\ref{m1n}) have the same column structure, reflecting the two physical charges of matter:
\begin{eqnarray}
j^{y_1\cdots y_n Y_0}  \!&=&\! \left(\begin{array}{c}j^{y_1\cdots y_n} \\ p^{y_1\cdots y_ny_0} \end{array}\right),\label{jMmax}\\
i^{y_1\cdots y_n Y_0y'}  \!&=&\! \left(\begin{array}{c}i^{y_1\cdots y_ny'} \\ k^{y_1\cdots y_ny_0y'} \end{array}\right),\label{iMmax}
\end{eqnarray}
whereas $m^{y_1\dots y_{n} \dot{Y}_0 } =0$.

As a result, the master equation (\ref{master}) reduces to the coupled system of the two
sets of equations for the moments:
\begin{eqnarray}
{\frac{D}{ds}} j^{y_1\cdots y_n}  &=& - n\, v^{(y_1} j^{y_2 \dots y_n)} + n\,i^{(y_1 \dots y_{n})},\label{dj1nmax}\\
{\frac{D}{ds}} p^{y_1\cdots y_n y_0}  &=& - n\, v^{(y_1} p^{y_2 \dots y_n)y_0} + n\,k^{(y_1 \dots y_{n-1}|y_0| y_{n})}\nonumber\\
&& - F_{y'}{}^{y_0}i^{y_1 \dots y_{n}y'} - \sum\limits^{\infty}_{k=1}{\frac 1{k!}}F_{y'}{}^{y_0}{}_{;y_{n+1}\dots y_{n+k}}i^{y_1 \dots y_{n+k}y'}.\label{dp1nmax}
\end{eqnarray}
These equations should be compared to those of \cite{Dixon:1967}.

\section{Equations of motion in metric-affine gravity}\label{eom_sec}

We are now in a position to derive the equations of motion for extended test bodies in metric-affine gravity. As a preliminary step, we rewrite the conservation laws (\ref{cons1b}) and (\ref{cons2b}) as \cite{Obukhov:Puetzfeld:2014}:
\begin{eqnarray}\label{cons1f}
\widetilde{\nabla}_j\Delta^i{}_k{}^j &=& -\,U_{jm}{}^{ni}{}_k\Delta^m{}_n{}^j + \Sigma_k{}^i - t_k{}^i,\\
\widetilde{\nabla}_j\Sigma_k{}^j &=& -\,V_{j}{}^n{}_k\Sigma_n{}^j - R_{kjm}{}^n\Delta^m{}_n{}^j - {\frac 12}Q_{kj}{}^nt_n{}^j - A_k\,L_{\rm mat}.\label{cons2f}
\end{eqnarray}
Here we denoted $A_k := \widetilde{\nabla}_k\log F$, and
\begin{eqnarray}
U_{jmn}{}^{ik} &:=& A_j\delta_m^i\delta_n^k - N_{jm}{}^i\delta_n^k + N_{j}{}^k{}_n\delta_m^i,\label{U}\\
V_{jn}{}^k &:=& A_j\delta_n^k + N^k{}_{jn}.\label{V}
\end{eqnarray}

Introducing the dynamical current
\begin{equation}
J^{Aj} = \left(\begin{array}{c}\Delta^{ikj} \\ \Sigma^{kj}\end{array}\right),\label{JAM}
\end{equation}
and the material current
\begin{equation}
K^{\dot{A}} = \left(\begin{array}{c} t^{ik} \\ L_{\rm mat}\end{array}\right),\label{XIM}
\end{equation}
we then recast the system (\ref{cons1f}) and (\ref{cons2f}) into the generic conservation law (\ref{dJ}), where we now have 
\begin{eqnarray}
\Lambda_{jB}{}^A &=& \left(\begin{array}{c|c}U_{ji'k'}{}^{ik} & - \delta^i_j\delta^k_{k'}\\ \hline R^k{}_{ji'k'} & V_{jk'}{}^k\end{array}\right),\label{LamMAG}\\
  \Pi^A{}_{\dot{B}} &=& \left(\begin{array}{c|c}\delta^i_{i'}\delta^k_{k'} & 0\\ \hline {\frac 12} Q^k{}_{i'k'} & A^k\end{array}\right).\label{PIMAG}
\end{eqnarray}

Like in the previous example of an electrically charged body, the matter elements in metric-affine gravity are also characterized by two ``charges'': the canonical hypermomentum (upper component) and the canonical energy-momentum (lower component). This is reflected in the column structure of the dynamical current (\ref{JAM}). The material current (\ref{XIM}) takes into account the metrical energy-momentum and the matter Lagrangian related to the nonminimal coupling. The multi-index $A = \{ik,k\}$, whereas $\dot{A} = \{ik,1\}$. Accordingly, the generalized propagator reads
\begin{equation}
G^Y{}_X = \left(\begin{array}{c|c} g^{y_1}{}_{x_1}g^{y_2}{}_{x_2} & 0\\ \hline 0 & g^{y_1}{}_{x_1}\end{array}\right),\label{propMAG}
\end{equation}
and we easily construct the expansion coefficients of its derivatives from the corresponding expansions of the derivatives of vector propagator $g^{y}{}_{x}$:
\begin{eqnarray}
\gamma^{Y_0}{}_{Y_1y_2\dots y_{k+2}} = \left(\begin{array}{c|c} \gamma^{\{y_0\tilde{y}\}}{}_{\{y'y''\}y_2\dots y_{k+2}} & 0\\ \hline 0 & \gamma^{y_0}{}_{y'y_2\dots y_{k+2}}\end{array}\right),\label{propE}
\end{eqnarray}
where we denoted 
\begin{equation}
\gamma^{\{y_0\tilde{y}\}}{}_{\{y'y''\}y_2\dots y_{k+2}} = \gamma^{y_0}{}_{y'y_2\dots y_{k+2}}\delta^{\tilde{y}}_{y''} +  \gamma^{\tilde{y}}{}_{y''y_2\dots y_{k+2}}\delta^{y_0}_{y'}.\label{ggg}
\end{equation}
In particular, for the first expansion coefficient ($k = 1$), we find
\begin{eqnarray}
\gamma^{\{y_0\tilde{y}\}}{}_{\{y'y''\}y_2y_{3}} &=& {\frac 12}\left(\widetilde{R}^{y_0}{}_{y'y_2y_3}\delta^{\tilde{y}}_{y''} + \widetilde{R}^{\tilde{y}}{}_{y''y_2y_3}\delta^{y_0}_{y'}\right),\label{gR1}\\
\gamma^{y_0}{}_{y'y_2y_{3}} &=& {\frac 12}\widetilde{R}^{y_0}{}_{y'y_2y_3}.\label{gR2}
\end{eqnarray}

For completeness, let us write down also another generalized propagator 
\begin{equation}
G^{\dot{Y}}{}_{\dot{X}} = \left(\begin{array}{c|c} g^{y_1}{}_{x_1}g^{y_2}{}_{x_2} & 0\\ \hline 0 & 1\end{array}\right).\label{propMAG2}
\end{equation}

The last step is to write the generalized moments (\ref{j1n})-(\ref{m1n}) in terms of their components:
\begin{eqnarray}
j^{y_1\cdots y_n Y}  \!&=&\! \left(\begin{array}{c}h^{y_1\cdots y_ny'y''} \\ p^{y_1\cdots y_ny'} \end{array}\right),\label{jMAG}\\
i^{y_1\dots y_{n} Y y_0} \!&=&\! \left(\begin{array}{c}q^{y_1\cdots y_ny'y''y_0} \\ k^{y_1\cdots y_ny'y_0} \end{array}\right),\label{iMAG}\\
m^{y_1\dots y_{n} \dot{Y} } \!&=&\! \left(\begin{array}{c}\mu^{y_1\cdots y_ny'y''} \\ \xi^{y_1\cdots y_n} \end{array}\right).\label{mMAG}
\end{eqnarray}
For the two most important moments, ``$h$'' stands for the hypermomentum, whereas ``$p$'' stands for the momentum.\footnote{Note that in order to facilitate the comparison with our previous work \cite{Puetzfeld:Obukhov:2013:3}, we provide in appendix the explicit form of integrated conservation laws (\ref{cons1f}) and (\ref{cons2f}), as well as the generalized integrated moments (\ref{jMAG}) -- (\ref{mMAG}) in the notation used in \cite{Puetzfeld:Obukhov:2013:3}.} Finally, substituting all the above into the ``master equation'' (\ref{master}), we obtain the system of multipolar equations of motion for extended test bodies in metric-affine gravity:
\begin{eqnarray}
\frac{D}{ds} h^{y_1 \dots y_n y_a y_b} &=&  - n \, v^{(y_1} h^{y_2 \dots y_n) y_a y_b} + n \, q^{(y_1 \dots y_{n-1} | y_a y_b | y_n)} \nonumber \\ 
&&+ k^{y_1 \dots y_n y_b y_a} - \mu^{y_1 \dots y_n y_a y_b}  \nonumber \\
&& - \frac{1}{2}\widetilde{R}^{y_a}{}_{y' y'' y_{n+1}} \left(q^{y_1 \dots y_{n+1} y' y_b y''} + v^{y''} h^{y_1 \dots y_{n+1} y' y_b}\right)  \nonumber \\
&& - \frac{1}{2}\widetilde{R}^{y_b}{}_{y' y'' y_{n+1}}\left( q^{y_1 \dots y_{n+1} y_a y' y''} + v^{y''} h^{y_1 \dots y_{n+1} y_a y'}\right)   \nonumber \\
&& - U_{y_0y'y''}{}^{y_ay_b} q^{y_1 \dots y_n y' y'' y_0} - U_{y_0y'y''}{}^{y_ay_b}{}_{;y_{n+1}} q^{y_1 \dots y_{n+1} y' y'' y_0}\nonumber\\
&& + \sum^{\infty}_{k=2} {\frac{1}{k!}}\Bigg[ (-1)^k  v^{y'} n \, \beta^{(y_1}{}_{y' y_{n+1} \dots y_{n+k}} h^{y_2 \dots y_n)y_{n+1} \dots y_{n+k} y_a y_b} \nonumber \\ 
&& +(-1)^k \gamma^{y_a}{}_{y' y'' y_{n+1} \dots y_{n+k}}\left( q^{y_1 \dots y_{n+k} y' y_b y''} + v^{y''} h^{y_1 \dots y_{n+k} y' y_b}\right) \nonumber \\
&& + (-1)^k \gamma^{y_b}{}_{y' y'' y_{n+1} \dots y_{n+k}}\left( q^{y_1 \dots y_{n+k} y_a y' y''} + v^{y''} h^{y_1 \dots y_{n+k} y_a y'}\right)   \nonumber \\
&& - (-1)^k n \, \alpha^{(y_1}{}_{y' y_{n+1} \dots y_{n+k}} q^{y_2 \dots y_n)y_{n+1} \dots y_{n+k} y_a y_b y'} \nonumber \\
&& -  U_{y_0y'y''}{}^{y_ay_b}{}_{;y_{n+1} \dots y_{n+k}} q^{y_1 \dots y_{n+k} y' y'' y_0} \Bigg], \label{int_eom_1_general}
\end{eqnarray}
\begin{eqnarray}
\frac{D}{ds} p^{y_1 \dots y_n y_a}&=& - n \, v^{(y_1} p^{y_2 \dots y_n) y_a} + n \, k^{(y_1 \dots y_{n-1} | y_a | y_n)} \nonumber \\ 
&& - \,A^{y_a} \xi^{y_1 \dots y_n} - A^{y_a}{}_{;y_{n+1}} \xi^{y_1 \dots y_{n+1}} \nonumber\\
&& - \,V_{y''y'}{}^{y_a} k^{y_1 \dots y_n y' y''} - V_{y''y'}{}^{y_a}{}_{;y_{n+1}} k^{y_1 \dots y_{n+1} y' y''} \nonumber \\
&& - \,\frac{1}{2} \widetilde{R}^{y_a}{}_{y' y'' y_{n+1}}\left(k^{y_1 \dots y_{n+1} y' y''} + v^{y''} p^{y_1 \dots y_{n+1} y'} \right)\nonumber \\
&&   -\,R^{y_a}{}_{y_0 y' y''} q^{y_1 \dots y_n y' y'' y_0} - R^{y_a}{}_{y_0 y' y'';y_{n+1}} q^{y_1 \dots y_{n+1} y' y'' y_0} \nonumber \\
&&  -\,\frac{1}{2} Q^{y_a}{}_{y'' y'} \mu^{y_1 \dots y_n y' y''}  - \frac{1}{2} Q^{y_a}{}_{y'' y';y_{n+1}} \mu^{y_1 \dots y_{n+1} y' y''} \nonumber \\
&& + \sum^{\infty}_{k=2} \frac{1}{k!}\Bigg[ (-1)^k n \, v^{y'} \beta^{(y_1}{}_{y' y_{n+1} \dots y_{n+k}} p^{y_2 \dots y_n ) y_{n+1} \dots y_{n+k} y_a} \nonumber \\ 
&&+(-1)^k \gamma^{y_a}{}_{y' y'' y_{n+1} \dots y_{n+k}}\left( k^{y_1 \dots y_{n+k} y' y''} +  v^{y''} p^{y_1 \dots y_{n+k} y'}\right) \nonumber \\ 
&&  -\,(-1)^k n \, \alpha^{(y_1}{}_{y' y_{n+1} \dots y_{n+k}} k^{y_2 \dots y_n)y_{n+1} \dots y_{n+k} y_a y' } \nonumber \\
&&  -\,R^{y_a}{}_{y_0 y' y'';y_{n+1} \dots y_{n+k}} q^{y_1 \dots y_{n+k} y' y'' y_0}\nonumber\\
&& -\,V_{y''y'}{}^{y_a}{}_{;y_{n+1} \dots y_{n+k}} k^{y_1 \dots y_{n+k} y' y''} \nonumber \\
&&  -\,\frac{1}{2} Q^{y_a}{}_{y'' y';y_{n+1} \dots y_{n+k}} \mu^{y_1 \dots y_{n+k} y' y''} \nonumber \\
&& -\,A^{y_a}{}_{;y_{n+1} \dots y_{n+k}} \xi^{y_1 \dots y_{n+k}}  \Bigg]. 
\label{int_eom_2_general}
\end{eqnarray}

\subsection{Special cases} \label{special_cases_sec}

The general equations of motion (\ref{int_eom_1_general}) and (\ref{int_eom_2_general}) are valid to {\it any} multipolar order. In the following sections we focus on some special cases, in particular we work out the two lowest multipolar orders of approximation, and consider the explicit form of the equations of motion in special geometries.

\subsubsection{General pole-dipole equations of motion}

From (\ref{int_eom_1_general}) and (\ref{int_eom_2_general}), we can derive the general pole-dipole equations of motion. The relevant moments to be kept at this order of approximation are: $p^a, p^{ab}, h^{ab}, q^{abc}, k^{ab}, k^{abc}, \mu^{ab}, \mu^{abc}, \xi^{a},$ and $\xi$. Since all objects are now evaluated on the world-line, we switch back to the usual tensor notation.

For $n=1$ and $n=0$, eq.\ (\ref{int_eom_1_general}) yields
\begin{eqnarray}
0 &=& k^{acb} - \mu^{abc} + q^{bca} - v^a h^{bc}, \label{eom_1_n_1} \\
\frac{D}{ds} h^{ab} &=&  k^{ba} - \mu^{ab} - U_{cde}{}^{ab} q^{dec}.  \label{eom_1_n_0}
\end{eqnarray}
Furthermore for $n=2,1,0$ equation (\ref{int_eom_2_general}) yields
\begin{eqnarray}
0 &=& k^{(a|c|b)} - v^{(a} p^{b)c}, \label{eom_2_n_2} \\
\frac{D}{ds} p^{ab} &=&  k^{ba} - v^a p^b  - A^b \xi^a  - V_{dc}{}^b k^{acd} - \frac{1}{2} Q^b{}_{dc} \mu^{acd},\label{eom_2_n_1} \\
\frac{D}{ds} p^{a} &=& - V_{cb}{}^a k^{bc} - R^{a}{}_{dbc} q^{bcd} - \frac{1}{2} Q^a{}_{cb} \mu^{bc} \nonumber \\
&& - A^a \xi -\frac{1}{2} \widetilde{R}^a{}_{cdb} \left(k^{bcd} + v^d p^{bc} \right) \nonumber \\
&&- V_{dc}{}^a{}_{;b} k^{bcd} - \frac{1}{2} Q^a{}_{dc;b} \mu^{bcd} - A^a{}_{;b} \xi^b. \label{eom_2_n_0} 
\end{eqnarray}

\paragraph{Rewriting the equations of motion}

Let us decompose (\ref{eom_1_n_1}) and (\ref{eom_1_n_0}) into symmetric and skew-symmetric parts:
\begin{eqnarray}
\mu^{abc} &=& k^{a(bc)} + q^{(bc)a} - v^a h^{(bc)}, \label{eom_1_n_1S} \\
0 &=& -\,k^{a[bc]} + q^{[bc]a} - v^a h^{[bc]}, \label{eom_1_n_1A} \\
\mu^{ab} &=& -\,\frac{D}{ds}h^{(ab)} +  k^{(ab)} - U_{cde}{}^{(ab)} q^{dec},  \label{eom_1_n_0S}\\
\frac{D}{ds} h^{[ab]} &=& -\,k^{[ab]} - U_{cde}{}^{[ab]} q^{dec}.  \label{eom_1_n_0A}
\end{eqnarray}
As a result, we can express the moments symmetric in the two last indices $\mu^{ab} = \mu^{(ab)}$ and $\mu^{cab} = \mu^{c(ab)}$ (in general, this is possible also for an arbitrary order $\mu^{c_1\dots c_nab} = \mu^{c_1\dots c_n(ab)}$) in terms of the other moments. 

Let us denote the skew-symmetric part $s^{ab} := h^{[ab]}$, as this simplifies greatly the subsequent manipulations and the comparison with  \cite{Puetzfeld:Obukhov:2013:3}.

The system of the two equations (\ref{eom_2_n_2}) and (\ref{eom_1_n_1A}) can be resolved in terms of the 3rd rank $k$-moment. The result reads explicitly
\begin{eqnarray}
k^{abc} &=& v^a p^{cb} + v^c\left(p^{[ab]} - s^{ab}\right) 
+ v^b\left(p^{[ac]} - s^{ac}\right) + v^a\left(p^{[bc]} - s^{bc}\right)\nonumber\\
&& + q^{[ab]c} + q^{[ac]b} + q^{[bc]a}.\label{kabc}
\end{eqnarray}
This yields some useful relations:
\begin{eqnarray}
k^{a[bc]} &=& - v^as^{bc} + q^{[bc]a},\label{ka1}\\
k^{[ab]c} &=&  v^{[a} p^{|c|b]} + v^c\left(p^{[ab]} - s^{ab}\right) + q^{[ab]c}.\label{ka2}
\end{eqnarray}

The next step is to use the equations (\ref{eom_1_n_1S}), (\ref{eom_1_n_0S}) together with (\ref{kabc}) and to substitute the $\mu$-moments and $k$-moments into (\ref{eom_1_n_0}) and (\ref{eom_2_n_1})-(\ref{eom_2_n_0}). This yields a system that depends only on the $p,h,q,$ and $\xi$ moments. 

Let us start with the analysis of (\ref{eom_2_n_0}). The latter contains the combination $k^{[b|c|d]} + v^{[d}p^{b]c}$ where the skew symmetry is imposed by the contraction with the Riemann curvature tensor, which is antisymmetric in the two last indices. Making use of (\ref{kabc}), we derive
\begin{equation}
k^{[a|c|b]} + v^{[b}p^{a]c} = \kappa^{abc} + \kappa^{acb} - \kappa^{bca},\label{kka}
\end{equation}
where we introduced the abbreviation
\begin{equation}
\kappa^{abc} := v^c\left(p^{[ab]} - s^{ab}\right) +  q^{[ab]c}.\label{kappa}
\end{equation}
Note that by construction $\kappa^{abc} = \kappa^{[ab]c}$. 

Then by making use of the Ricci identity we find 
\begin{eqnarray}
-\,{\frac 12} \widetilde{R}^a{}_{cdb} \left(k^{bcd} + v^d p^{bc} \right) &=& 
\widetilde{R}^a{}_{bcd}\left[q^{[cd]b}  + v^b\left(p^{[cd]} - s^{cd}\right)\right].\label{Rq}
\end{eqnarray}
Substituting $k^{bc}$ from (\ref{eom_2_n_1}) and $\mu^{bc}$ from (\ref{eom_1_n_0S}), we find after some algebra
\begin{eqnarray}
&&-\,V_{cb}{}^ak^{bc} - {\frac 12}Q^a{}_{cb}\mu^{bc} = - \,A_b{\frac {Dp^{ba}}{ds}} - N^a{}_{cd}{\frac {Dh^{cd}}{ds}} - \left(p^a + N^a{}_{cd}h^{cd}\right)v^bA_b \nonumber\\
&& - A^aA^b\xi_b - k^{bac}A_bA_c + \left(N^a{}_{nb}N_{dc}{}^n - N^a{}_{cn}N_d{}^n{}_b\right)q^{cbd}.\label{VkQmu1}
\end{eqnarray}
Further simplification is achieved by noticing that
\begin{eqnarray}
v^bA_b = {\frac {DA}{ds}},\qquad
k^{bac}A_bA_c = p^{ca}A_c{\frac {DA}{ds}},\label{kAA}
\end{eqnarray}
where we used (\ref{eom_2_n_2}) and recalled that $A_b = A_{;b}$.

Analogously, taking $k^{b[cd]}$ from (\ref{ka1}) and $\mu^{bcd}$ from (\ref{eom_1_n_1S}), we derive
\begin{eqnarray}
-\,V_{dc}{}^a{}_{;b}k^{bcd} - {\frac 12}Q^a{}_{dc;b}\mu^{bcd} = - \,A_{b;c}k^{cab}+ N^a{}_{cd;b}q^{cdb} - N^a{}_{cd;b}v^bh^{cd}.\label{VkQmu2}
\end{eqnarray}
We can again use $A_b = A_{;b}$ and (\ref{eom_2_n_2}) to simplify
\begin{equation}
- \,A_{b;c}k^{cab} = - p^{ba}{\frac {DA_b}{ds}}.\label{kAd}
\end{equation}

After these preliminary calculations, we substitute (\ref{Rq})-(\ref{kAd}) into (\ref{eom_2_n_0}) to recast the latter into 
\begin{eqnarray}
&&{\frac {D}{ds}}\left(Fp^a + FN^a{}_{cd}h^{cd} + p^{ba}\widehat{\nabla}_bF\right)  = F \widetilde{R}^a{}_{bcd}v^b\left(p^{[cd]} - s^{cd}\right) \nonumber \\
&&- Fq^{cbd}\left[R^a{}_{dcb} - \widetilde{R}^a{}_{dcb} - N^a{}_{cb;d} - N^a{}_{nb}N_{dc}{}^n + N^a{}_{cn}N_d{}^n{}_b\right]\nonumber\\
&& - FA^a\left(\xi + \xi^bA_b\right) - F\xi^bA^a{}_{;b}.\label{DPa}
\end{eqnarray}

Finally, combining (\ref{eom_1_n_0}) and (\ref{eom_2_n_1}) to eliminate $k^{ba}$ we derive the equation
\begin{eqnarray}
{\frac {D}{ds}}\left(p^{ab} - h^{ab}\right) &=& \mu^{ab} - v^a\left(p^b + N^b{}_{cd}h^{cd}\right)\nonumber\\
&& +\,q^{cda}N^b{}_{cd} - q^{cbd}N_{dc}{}^a + q^{acd}N_d{}^b{}_c\nonumber\\
&& -\,\xi^aA^b + (q^{abc} - k^{abc})A_c.\label{Dpab}
\end{eqnarray}

Following \cite{Puetzfeld:Obukhov:2013:3}, we introduce the total orbital and the total spin angular moments
\begin{equation}
L^{ab} := 2p^{[ab]},\qquad S^{ab} := -2h^{[ab]},\label{LS}
\end{equation}
and define the generalized total energy-momentum 4-vector and the generalized total angular momentum by
\begin{eqnarray}
{\cal P}^a &:=& F(p^a + N^a{}_{cd}h^{cd}) + p^{ba}\widehat{\nabla}_bF,\label{Ptot}\\
{\cal J}^{ab} &:=& F(L^{ab} + S^{ab}).\label{Jtot}
\end{eqnarray}
Then, taking into account the identity (\ref{RRN}), which with the help of the raising and lowering of indices can be recast into
\begin{eqnarray}
\widetilde{\nabla}^aN_{dcb} = - R^a{}_{dcb} + \widetilde{R}^a{}_{dcb} + N^a{}_{cb;d} 
+ N^a{}_{nb}N_{dc}{}^n - N_d{}^n{}_bN^a{}_{cn},\label{DNRR}
\end{eqnarray}
we rewrite the pole-dipole equations of motion (\ref{DPa}) and (\ref{Dpab}) in the final form
\begin{eqnarray}
{\frac {D{\cal P}^a}{ds}} &=& {\frac 12}\widetilde{R}^a{}_{bcd}v^b{\cal J}^{cd} + Fq^{cbd}\widetilde{\nabla}^aN_{dcb} \nonumber\\ 
&& -\, \xi\widetilde{\nabla}^aF - \xi^b\widetilde{\nabla}_b\widetilde{\nabla}^aF,\label{DPtot}\\
{\frac {D{\cal J}^{ab}}{ds}} &=& -\,2v^{[a}{\cal P}^{b]} + 2F(q^{cd[a}N^{b]}{}_{cd} + q^{c[a|d|}N_{dc}{}^{b]} + q^{[a|cd|}N_d{}^{b]}{}_c)\nonumber\\
&& - 2\xi^{[a}\widetilde{\nabla}^{b]}F.\label{DJtot}
\end{eqnarray}
The last equation arises as the skew-symmetric part of (\ref{Dpab}), whereas the symmetric part of the latter is a non-dynamical relation that determines the $\mu^{ab}$ moment
\begin{eqnarray}
\mu^{ab} &=& {\frac {D\Upsilon^{ab}}{ds}} + {\frac 1F} v^{(a} \left( {\cal P}^{b)} + {\cal J}^{b)c}A_c \right) + \xi^{(a}A^{b)}\nonumber\\
&& -\,q^{cd(a}N^{b)}{}_{cd} + q^{c(a|d|}N_{dc}{}^{b)} - q^{(a|cd|}N_d{}^{b)}{}_c\nonumber\\ 
&& +\,(q^{[ac]b} + q^{[bc]a} - q^{(ab)c})A_c.\label{muabPD}
\end{eqnarray}
Here the symmetric moment of the total hypermomentum is introduced via
\begin{equation}
\Upsilon^{ab} := p^{(ab)} - h^{(ab)}.\label{hypermom}
\end{equation}

\subsubsection{Conserved quantity}\label{cons_mom}

The equations of motion for the multipole moments are derived from the conservation laws of the energy-momentum and the hypermomentum currents $\Sigma_k{}^i$ and $\Delta^m{}_n{}^i$. In section \ref{Induced_subsec} we demonstrated that every generalized Killing vector induces a conserved current constructed from $\Sigma_k{}^i$ and $\Delta^m{}_n{}^i$. Quite remarkably, there is a direct counterpart of such an induced current built from the multipole moments. 

Let $\zeta^k$ be a generalized Killing vector, and let us contract equation (\ref{DPtot}) with $\zeta_a$ and equation (\ref{DJtot}) with ${\frac 12}\widetilde{\nabla}_a\zeta_b$, and then take the sum. This yields
\begin{equation}\label{Dlie}
{\frac {D}{ds}}\left({\cal P}^a\zeta_a + {\frac 12}{\cal J}^{ab}\widetilde{\nabla}_a\zeta_b\right) = Fq^{cbd}{\cal L}_\zeta N_{dcb} - \xi\,{\cal L}_\zeta F - \xi^a\widetilde{\nabla}_a{\cal L}_\zeta F.
\end{equation}
On the right-hand side, the Lie derivatives of the distortion tensor and of the coupling function both vanish in view of (\ref{LieN}) and (\ref{LieF}).  

Consequently, we conclude that for every generalized Killing vector field the quantity
\begin{equation}\label{consPJ}
{\cal P}^a\zeta_a + {\frac 12}{\cal J}^{ab}\widetilde{\nabla}_a\zeta_b = {\rm const}
\end{equation}
is conserved along the trajectory of an extended body. 

We thus observe a complete consistency between (\ref{DI1}), (\ref{I1}), (\ref{I2})  and (\ref{consPJ}), (\ref{Dlie}). 

\subsubsection{Coupling to the post-Riemannian geometry: Fine structure}

Let us look more carefully at how the post-Riemannian pieces of the gravitational field couple to extended test bodies. At first, we notice that the generalized energy-momentum vector (\ref{Ptot}) contains the term $N^a{}_{cd}h^{cd}$ that describes the direct interaction of the distortion (torsion plus nonmetricity) with the intrinsic dipole moment of the hypermomentum. Decomposing the latter into the skew-symmetric (spin) part and the symmetric (proper hypermomentum + dilation) part, we find
\begin{equation}
N^a{}_{cd}h^{cd} = -\,{\frac 12}N^a{}_{[cd]}S^{cd} - {\frac 12}Q^a{}_{cd}h^{(cd)}.\label{Nh}
\end{equation}
Here we made use of (\ref{QN}). This is well consistent with the gauge-theoretic structure of metric-affine gravity. The second term shows that the intrinsic proper hypermomentum and the dilation moment couple to the nonmetricity, whereas the first term displays the typical spin-torsion coupling. 

Similar observations can be made for the coupling of higher moments which appear on the right-hand sides of (\ref{DPtot}) and (\ref{DJtot}) - and thus determine the force and torque acting on an extended body due to the post-Riemannian gravitational field. In order to see this, let us introduce the decomposition 
\begin{equation}
-\,{\frac 12}q^{abc} = {\stackrel {d}{q}}{}^{abc} + {\stackrel {s}{q}}{}^{cab}\label{qqq}
\end{equation}
into the two pieces 
\begin{eqnarray}
{\stackrel {d}{q}}{}^{abc} &:=& {\frac 12}\left(q^{[ac]b} + q^{[bc]a} - q^{(ab)c}\right),\label{qd}\\
{\stackrel {s}{q}}{}^{abc} &:=& {\frac 12}\left(q^{[ab]c} + q^{[ac]b} - q^{[bc]a}\right).\label{qs}
\end{eqnarray}
The overscript ``$d$'' and ``$s$'' notation shows the relevance of these objects to the dilation plus proper hypermomentum and to the spin, respectively. By construction, we have the following algebraic properties
\begin{equation}
{\stackrel {d}{q}}{}^{[ab]c} \equiv 0,\qquad {\stackrel {s}{q}}{}^{(ab)c}  \equiv 0.\label{qq0}
\end{equation}

Making use of the decomposition (\ref{qqq}) and of the explicit structure of the distortion (\ref{NTQ}), we then recast the equations of motion (\ref{DPtot}) and (\ref{DJtot}) into
\begin{eqnarray}
{\frac {D{\cal P}^a}{ds}} &=& {\frac 12}\widetilde{R}^a{}_{bcd}v^b{\cal J}^{cd}\nonumber\\ 
&& +\,F{\stackrel {s}{q}}{}^{cbd}\widetilde{\nabla}^a T_{cbd} + F{\stackrel {d}{q}}{}^{cbd}\widetilde{\nabla}^a Q_{dcb}\nonumber\\ 
&& -\,\xi\widetilde{\nabla}^aF - \xi^b\widetilde{\nabla}_b\widetilde{\nabla}^aF,\label{DPdec}\\
{\frac {D{\cal J}^{ab}}{ds}} &=& -\,2v^{[a}{\cal P}^{b]}\nonumber\\ 
&& +\,2F({\stackrel {s}{q}}{}^{cd[a}T_{cd}{}^{b]} + 2{\stackrel {s}{q}}{}^{[a|cd|}T^{b]}{}_{cd})\nonumber\\
&& +\, 2F({\stackrel {d}{q}}{}^{cd[a}Q^{b]}{}_{cd} + 2{\stackrel {d}{q}}{}^{[a|dc|}Q_{cd}{}^{b]})\nonumber\\ 
&& -\,2\xi^{[a}\widetilde{\nabla}^{b]}F.\label{DJdec}
\end{eqnarray}
Now we clearly see the fine structure of the coupling of extended bodies to the post-Riemannian geometry. The first lines in the equations of motion describe the usual Mathisson-Papapetrou force and torque. They depend on the Riemannian geometry only. A body with the nontrivial moment (\ref{qs}) is affected by the torsion field, whereas the nontrivial moment (\ref{qd}) feels the nonmetricity. This explains the different physical meaning of the higher moments (\ref{qd}) and (\ref{qs}). In addition, the last lines in (\ref{DPdec}) and (\ref{DJdec}) describe contributions due to the nonminimal coupling. 

\subsubsection{General monopolar equations of motion}

At the monopolar order we have nontrivial moments $p^a, k^{ab}, \mu^{ab}$ and $\xi$. The nontrivial equations of motion then arise from (\ref{int_eom_1_general}) for $n = 0$ and from (\ref{int_eom_2_general}) for $n=1, n = 0$:
\begin{eqnarray}
0 &=& k^{ba} - \mu^{ab},\label{Mono1}\\
0 &=& k^{ba} - v^a p^b,\label{Mono2}\\
{\frac {Dp^a}{ds}} &=& -\,V_{cb}{}^ak^{bc} - {\frac 12}Q^a{}_{cb}\mu^{bc} - A^a\xi.\label{Mono3}
\end{eqnarray}
The first two equations (\ref{Mono1}) and (\ref{Mono2}) yield
\begin{equation}
k^{[ab]} = 0,\qquad v^{[a} p^{b]} = 0,\label{vp}
\end{equation}
and substituting (\ref{Mono1}), (\ref{Mono2}) and (\ref{vp}) into (\ref{Mono3}) we find
\begin{equation}
{\frac {D(Fp^a)}{ds}} = -\,\xi\widetilde{\nabla}^aF.\label{Mono4}
\end{equation}
{}From (\ref{vp}) we have $p^a = Mv^a$ with the mass $M := v^ap_a$, and this allows to recast (\ref{Mono4}) into the final form
\begin{equation}
M{\frac {Dv^a}{ds}} = -\,\xi(g^{ab} - v^av^b){\frac {\widetilde{\nabla}_bF}F}.\label{Mono5}
\end{equation}
Hence, in general the motion of nonminimally coupled monopole test bodies is nongeodetic. Furthermore, the general monopole equation of motion (\ref{Mono5}) reveals an interesting feature of theories with nonminimal coupling. There is an ``indirect'' coupling, i.e.\ through the coupling function $F(g_{ij}, R_{ijk}{}^l,T_{ij}{}^k,Q_{ij}{}^k)$, of post-Riemannian spacetime features to structureless test bodies.  

\subsubsection{Weyl-Cartan spacetime}

In Weyl-Cartan spacetime the nonmetricity reads $Q_{kij} = Q_kg_{ij}$, where $Q_k$ is the Weyl covector. Hence the distortion is given by
\begin{eqnarray}\label{wc_distortion}
N_{kj}{}^i = K_{kj}{}^i + {\frac{1}{2}}\left(Q^i g_{kj} - Q_k\delta^i_j - Q_j\delta^i_k\right).
\end{eqnarray}
The contortion tensor is constructed from the torsion,
\begin{equation}
K_{kj}{}^i = -\,{\frac 12}(T_{kj}{}^i + T^i{}_{kj} + T^i{}_{jk}).\label{KT}
\end{equation}
As a result, the generalized momentum (\ref{Ptot}) in Weyl-Cartan spacetime takes the form:
\begin{eqnarray}
{\cal P}^a = Fp^a - {\frac F2}\left(K^a{}_{cd} S^{cd} - Q_bS^{ba} + Q^aD\right) + p^{ba}\widetilde{\nabla}_bF. \label{wc_ptot}
\end{eqnarray}
Here we introduced the {\it intrinsic dilation} moment $D := g_{ab}h^{ab}$. 

Substituting the distortion (\ref{wc_distortion}) into (\ref{DPtot}) and (\ref{DJtot}), we find the pole-dipole equations of motion in Weyl-Cartan spacetime:
\begin{eqnarray}
{\frac {D{\cal P}^a}{ds}} &=& {\frac 12}\widetilde{R}^a{}_{bcd}v^b{\cal J}^{cd} + F{\stackrel {s}{q}}{}^{cbd}\widetilde{\nabla}^a T_{cbd} \nonumber\\ 
&& +\,Z^b\widetilde{\nabla}^aQ_b -\xi\widetilde{\nabla}^aF - \xi^b\widetilde{\nabla}_b\widetilde{\nabla}^aF,\label{DPWC}\\
{\frac {D{\cal J}^{ab}}{ds}} &=& -\,2v^{[a}{\cal P}^{b]} + 2F({\stackrel {s}{q}}{}^{cd[a}T_{cd}{}^{b]} + 2{\stackrel {s}{q}}{}^{[a|cd|}T^{b]}{}_{cd})\nonumber\\
&& + 2FZ^{[a}Q^{b]} - 2\xi^{[a}\widetilde{\nabla}^{b]}F.\label{DJWC}
\end{eqnarray}
Here we introduced the trace of the modified moment (\ref{qd})
\begin{eqnarray}
Z^a := g_{bc}{\stackrel {d}{q}}{}^{bca} = {\frac 12}g_{bc}\left(q^{bac} - q^{bca} - q^{abc}\right).\label{Za}
\end{eqnarray}
It is coupled to the Weyl nonmetricity.

\subsubsection{Weyl spacetime}

Weyl spacetime \cite{Weyl:1923} is obtained as a special case of the results above for vanishing torsion. Hence the contortion is trivial 
\begin{eqnarray}
K_{abc} = 0. \label{contortion}
\end{eqnarray}
Taking this into account, the generalized momentum (\ref{wc_ptot}) and the equations of motion (\ref{DPWC}) and (\ref{DJWC}) are simplified even further.

It is interesting to note that besides a direct coupling of the dilation moment to the Weyl nonmetricity on the right-hand sides of (\ref{DPWC}) and (\ref{DJWC}), there is also a nontrivial coupling of the spin to the nonmetricity in (\ref{wc_ptot}). 

\subsubsection{Riemann-Cartan spacetime}

Another special case is obtained when the Weyl vector vanishes $Q_a = 0$. Equations (\ref{wc_ptot})-(\ref{DJWC}) then reproduce {\it in a covariant way} the findings of Yasskin and Stoeger \cite{Stoeger:Yasskin:1980} when the coupling is minimal ($F=1$). For nonminimal coupling we recover our earlier results in \cite{Puetzfeld:Obukhov:2013:3}. 

\section{Equations of motion in scalar-tensor theories}\label{scalar_tensor_eom_sec}

The geometrical arena of scalar-tensor theories is the Riemannian spacetime, hence $N_{kj}{}^i = 0$ which means that both the torsion $T_{ij}{}^k = 0$ and the nonmetricity $Q_{kij} = 0$ vanish. 

Scalar-tensor theories have a long history and they belong to the most straightforward generalizations of Einstein's general relativity theory. In the so-called Brans-Dicke theory \cite{Brans:Dicke:1961,Brans:1962:1,Brans:1962:2,Dicke:1962:1,Dicke:1964} a scalar field is introduced as a variable ``gravitational coupling constant'' (which is thus more correctly called a ``gravitational coupling function''). Similar formalisms were developed earlier by Jordan \cite{Jordan:1955,Jordan:1959}, Thiry \cite{Thiry:1951} and their collaborators using the 5-dimensional Kaluza-Klein approach. An overview of the history and developments of scalar-tensor theories can be found in \cite{Fujii:Maeda:2003,Brans:2005,Goenner:2012,Sotiriou:2014}.

Surprisingly little attention was paid to the equations of motion of extended test bodies in scalar-tensor theories. Some early discussions can be found in \cite{Brans:1962:2,Bergmann:1968,Wagoner:1970}, and in \cite{Damour:etal:1992} the dynamics of compact bodies was thoroughly studied in the framework of the post-Newtonian formalism. 

\subsection{Scalar-tensor theory: Jordan frame}

We consider the class of scalar-tensor theories along the lines of \cite{Damour:etal:1992} where the action $I = \int d^4x\,{\stackrel J{\mathfrak L}}$ is constructed on the manifold with the spacetime metric ${\stackrel Jg}{}_{ij}$. The Lagrangian density ${\stackrel {J}{\mathfrak L}} = {\stackrel J {\sqrt{-g}}}{\stackrel JL}$ has the following form
\begin{eqnarray}
{\stackrel JL} = {\frac 1{2\kappa}}\biggl(- F^2\widetilde{R}({\stackrel Jg}) + {\stackrel Jg}{}^{ij}{\stackrel J\gamma}{}_{AB}\partial_i\varphi^A\partial_j\varphi^B - 2{\stackrel JU}\biggr) 
+ L_{\rm m}(\psi,\partial\psi,{\stackrel Jg}{}_{ij}).\label{LJ}
\end{eqnarray}
This action is an extension of standard Brans-Dicke theory \cite{Will:1993} to the case when we have a multiplet of scalar fields $\varphi^A$ (capital Latin indices $A,B,C = 1,\dots, N$ label the components of the multiplet). Here $\kappa = 8\pi G/c^4$ denotes Einstein's gravitational constant and in general we have several functions of scalar fields,
\begin{equation}\label{AU}
F = F(\varphi^A),\qquad {\stackrel JU} = {\stackrel JU}(\varphi^A),\qquad {\stackrel J\gamma}_{AB} = {\stackrel J\gamma}{}_{AB}(\varphi^A).
\end{equation}
The Lagrangian $L_{\rm m}(\psi,\partial\psi,{\stackrel Jg}{}_{ij})$ depends on the matter fields $\psi$ and the gravitational field. 

The metric ${\stackrel Jg}{}_{ij}$ determines angles and intervals in the {\it Jordan reference frame}. The Riemannian curvature scalar $\widetilde{R}({\stackrel Jg})$ is constructed from the Jordan metric. With the help of the conformal transformation
\begin{equation}
{\stackrel Jg}{}_{ij}\longrightarrow g_{ij} = F^2{\stackrel Jg}{}_{ij}\label{ggt} 
\end{equation}
we obtain a different metric on the spacetime manifold. This is called an Einstein reference frame. 

\subsection{Conservation laws: Einstein frame}

In the {\it Einstein reference frame} the Lagrangian density in the scalar-tensor theory reads ${\mathfrak L} = \sqrt{-g}L$ with
\begin{eqnarray}
L = {\frac 1{2\kappa}}\left(- \widetilde{R} + g^{ij}\gamma_{AB}\partial_i\varphi^A\partial_j\varphi^B - 2U\right) 
+ {\frac 1{F^4}}L_{\rm mat}(\psi,\partial\psi,F^{-2}g_{ij}).\label{LE}
\end{eqnarray}
Here the scalar curvature $\widetilde{R}(g)$ is constructed from the Einstein metric $g_{ij}$, and
\begin{equation}
\gamma_{AB} = {\frac 1{F^2}}({\stackrel J\gamma}{}_{AB} + 6F_{,A}F_{,B}),\qquad 
U = {\frac 1{F^4}}{\stackrel JU}.\label{gamUt}
\end{equation}

The metrical energy-mom\-entum tensor for the matter Lagrangian $L_{\rm mat}$ is constructed as usual via (\ref{tmet}). From the Noether theorem we find that it satisfies the generalized conservation law
\begin{equation}
\widetilde{\nabla}^it_{ij} = {\frac 1F}\left(4t_{ij} - g_{ij}t\right)\partial^iF.\label{dtt} 
\end{equation}
Here $t := g^{ij}t_{ij}$. The derivation is given in \cite{Obukhov:Puetzfeld:2014:2}. 

To begin with, we recast (\ref{dtt}) into an equivalent form 
\begin{equation}
\widetilde{\nabla}_it^{ij} = - A_i\left(\Xi^{ij} + t^{ij}\right)\label{cons}
\end{equation}
by introducing  
\begin{equation}
A_i := \partial_i\log F^{-4},\qquad \Xi^{ij} := - g^{ij}t/4.\label{AZ}
\end{equation}

\subsection{General equations of motion: arbitrary multipole order}

We now derive the equations of motion for extended test bodies in scalar-tensor gravity by making use of the master equations obtained for the general case of MAG. In scalar-tensor theory, the hypermomentum is zero $\Delta^{ikj} = 0$. Following the general scheme, we introduce the dynamical current which is a special case of the MAG current:
\begin{equation}
J^{Aj} = \left(\begin{array}{c}0 \\ \Sigma^{kj}\end{array}\right).\label{JAST}
\end{equation}
Taking into account the structure of the conservation laws (\ref{cons}) and (\ref{AZ}), we define the material current
\begin{equation}
K^{\dot{A}} = \left(\begin{array}{c} t^{ik} \\ -\,t/4\end{array}\right),\label{XIST}
\end{equation}
we then recast the system (\ref{cons}) into the generic conservation law, where we now have 
\begin{eqnarray}
\Lambda_{jB}{}^A &=& \left(\begin{array}{c|c} 0 & - \delta^i_j\delta^k_{k'}\\ \hline 0 & V_{jk'}{}^k\end{array}\right),\label{LamST}\\
\Pi^A{}_{\dot{B}} &=& \left(\begin{array}{c|c}\delta^i_{i'}\delta^k_{k'} & 0\\ \hline 0 & A^k\end{array}\right).\label{PIST}
\end{eqnarray}
In accordance with (\ref{cons}) and (\ref{AZ}) we now have
\begin{equation}
V_{jn}{}^k = A_j\delta_n^k,\qquad A_i = \partial_i\log F^{-4}.\label{VA}
\end{equation}
In view of (\ref{JAST})-(\ref{XIST}) the generalized moments read
\begin{eqnarray}
j^{y_1\cdots y_n Y}  \!&=&\! \left(\begin{array}{c} 0 \\ p^{y_1\cdots y_ny'} \end{array}\right),\label{jST}\\
i^{y_1\dots y_{n} Y y_0} \!&=&\! \left(\begin{array}{c} 0 \\ k^{y_1\cdots y_ny'y_0} \end{array}\right),\label{iST}\\
m^{y_1\dots y_{n} \dot{Y} } \!&=&\! \left(\begin{array}{c}\mu^{y_1\cdots y_ny'y''} \\ \xi^{y_1\cdots y_n} \end{array}\right).\label{mST}
\end{eqnarray}
It is worthwhile to notice that although we formally use a different notation for the upper and lower components of the moment (\ref{mST}), they are not independent. Recalling (\ref{XIST}), we have the obvious relation
\begin{equation}
\xi^{y_1\dots y_n} = - \,{\frac 14}g_{y'y''}\mu^{y_1\dots y_ny'y''},\label{ximu}
\end{equation}
which we will take into account when rewriting the equations of motion. 

Now we can make use of the general MAG equations of motion (\ref{master}). The first equation is a degenerate version of (\ref{int_eom_1_general}) since we have $h^{y_1 \dots y_n y_a y_b} = 0$ and $q^{y_1 \dots y_{n-1} y_a y_b y_n} = 0$, and we are left with the algebraic relation
\begin{equation}
k^{y_1 \dots y_n y_b y_a} = \mu^{y_1 \dots y_n y_a y_b}.\label{kmu}
\end{equation}
for all $n$. Hence we find 
\begin{equation}
\xi^{y_1\dots y_n} = - \,{\frac 14}g_{y'y''}k^{y_1\dots y_n y'y''},\label{xika}
\end{equation}
and the second equation (\ref{int_eom_2_general}) then gives rise to the equations of motion in scalar-tensor theory 
\begin{eqnarray}
\frac{D}{ds} p^{y_1 \dots y_n y_a}&=& - \,n \, v^{(y_1} p^{y_2 \dots y_n) y_a} + n \, k^{(y_1 \dots y_{n-1} | y_a | y_n)} \nonumber\\
&&- A^{y_a} \xi^{y_1 \dots y_n} - A^{y_a}{}_{;y_{n+1}} \xi^{y_1 \dots y_{n+1}} \nonumber\\
&&  - \,V_{y''y'}{}^{y_a} k^{y_1 \dots y_n y' y''} - V_{y''y'}{}^{y_a}{}_{;y_{n+1}} k^{y_1 \dots y_{n+1} y' y''} \nonumber\\
&& - \frac{1}{2} \widetilde{R}^{y_a}{}_{y' y'' y_{n+1}}\left(k^{y_1 \dots y_{n+1} y' y''} + v^{y''} p^{y_1 \dots y_{n+1} y'} \right)\nonumber \\
&& + \,\sum^{\infty}_{k=2} \frac{1}{k!}\Bigg[  - \,(-1)^k n \, \alpha^{(y_1}{}_{y' y_{n+1} \dots y_{n+k}} k^{y_2 \dots y_n)y_{n+1} \dots y_{n+k} y_a y' } \nonumber\\
&&+ (-1)^k n \, v^{y'} \beta^{(y_1}{}_{y' y_{n+1} \dots y_{n+k}} p^{y_2 \dots y_n ) y_{n+1} \dots y_{n+k} y_a} \nonumber \\
&& + (-1)^k \gamma^{y_a}{}_{y' y'' y_{n+1} \dots y_{n+k}}\left( k^{y_1 \dots y_{n+k} y' y''} +  v^{y''} p^{y_1 \dots y_{n+k} y'}\right)  \nonumber \\ 
&&  - \,V_{y''y'}{}^{y_a}{}_{;y_{n+1} \dots y_{n+k}} k^{y_1 \dots y_{n+k} y' y''} - A^{y_a}{}_{;y_{n+1} \dots y_{n+k}} \xi^{y_1 \dots y_{n+k}}  \Bigg]. \nonumber \\ \label{int_eom_ST}
\end{eqnarray}
The system (\ref{int_eom_ST}) is valid up to any multipole order. In the following we specialize it to the dipole and the monopole case.

\subsection{Pole-dipole equations of motion in scalar-tensor theory}

For $n=2,1,0$ equation (\ref{int_eom_ST}) yields
\begin{eqnarray}
0 &=& k^{(a|c|b)} - v^{(a} p^{b)c}, \label{eom_2_n_2_ST} \\
\frac{D}{ds} p^{ab} &=&  k^{ba} - v^a p^b  - A^b \xi^a  - V_{dc}{}^b k^{acd},\label{eom_2_n_1_ST}\\ 
\frac{D}{ds} p^{a} &=& - V_{cb}{}^a k^{bc} - V_{dc}{}^a{}_{;b} k^{bcd} - A^a{}_{;b} \xi^b \nonumber \\
&& - A^a \xi -\frac{1}{2} \widetilde{R}^a{}_{cdb} \left(k^{bcd} + v^d p^{bc} \right).
\label{eom_2_n_0_ST} 
\end{eqnarray}
Taking into account that $k^{a[bc]} = 0$, we resolve (\ref{eom_2_n_2_ST}) in a standard way to find explicitly
\begin{eqnarray}
k^{abc} = v^a p^{cb} + v^cp^{[ab]} + v^bp^{[ac]} + v^ap^{[bc]}.\label{kabcST}
\end{eqnarray}
In view of (\ref{xika}), we have in addition
\begin{eqnarray}
\xi^a &=& -\,{\frac 14}g_{bc}k^{abc},\label{xika1}\\
\xi &=& -\,{\frac 14}g_{ab}k^{ab}.\label{xika2}
\end{eqnarray} 
Then repeating the same algebra as we did in MAG, we recast the system (\ref{eom_2_n_1_ST}) and (\ref{eom_2_n_0_ST}) into 
\begin{eqnarray}
{\frac {D{\cal P}^a}{ds}} &=& {\frac 12}\widetilde{R}^a{}_{bcd}v^b{\cal J}^{cd} - \xi\widetilde{\nabla}^aF^{-4} - \xi^b\widetilde{\nabla}_b\widetilde{\nabla}^aF^{-4}, \label{DPtotST}\\
{\frac {D{\cal J}^{ab}}{ds}} &=& -\,2v^{[a}{\cal P}^{b]} - 2\xi^{[a}\widetilde{\nabla}^{b]}F^{-4}.\label{DJtotST}
\end{eqnarray}
Here, following \cite{Puetzfeld:Obukhov:2013:3}, we have the total orbital angular moment $
L^{ab} := 2p^{[ab]}$. Whereas the generalized total energy-momentum 4-vector and the generalized total angular momentum are introduced by
\begin{eqnarray}
{\cal P}^a &:=& F^{-4}p^a + p^{ba}\widetilde{\nabla}_bF^{-4},\label{PtotST}\\
{\cal J}^{ab} &:=& F^{-4}L^{ab}.\label{JtotST}
\end{eqnarray}

\subsection{Monopolar equations of motion in scalar-tensor theory}

At the monopolar order we find from eq.\ (\ref{int_eom_ST}) for $n=1, n = 0$:
\begin{eqnarray}
0 &=& k^{ba} - v^a p^b,\label{Mono1ST}\\
{\frac {Dp^a}{ds}} &=& -\,V_{cb}{}^ak^{bc} - A^a\xi.\label{Mono2ST}
\end{eqnarray}
Making use of $k^{[ab]} = 0$, the first equation yields $v^{[a} p^{b]} = 0$, hence we have 
\begin{equation}
p^a = Mv^a\label{pMv}
\end{equation}
with the mass $M := v^ap_a$. Substituting (\ref{VA}), (\ref{Mono1ST}) and (\ref{pMv}) into (\ref{Mono2ST}) we find
\begin{equation}
{\frac {D(F^{-4}Mv^a)}{ds}} = -\,\xi\widetilde{\nabla}^aF^{-4}.\label{Mono3ST}
\end{equation}
Contracting this with $v_a$, we derive 
\begin{equation}
{\frac {D(F^{-4}M)}{ds}} = -\,\xi v^a\widetilde{\nabla}_aF^{-4},\label{MFdot}
\end{equation}
we write (\ref{Mono3ST}) in the final form
\begin{equation}\label{Mono4ST}
M{\frac {Dv^a}{ds}} = -\,\xi(g^{ab} - v^av^b){\frac {\widetilde{\nabla}_bF^{-4}}{F^{-4}}}.
\end{equation}
Combining (\ref{xika2}) with (\ref{Mono1ST}) we find  
\begin{equation}
\xi = -\,{\frac {v^ap_a}4} = - \,{\frac M4}.\label{xiM} 
\end{equation}
Substituting this into (\ref{Mono4ST}), we obtain 
\begin{equation}\label{MonoF}
{\frac {Dv^a}{ds}} = -\,(g^{ab} - v^av^b){\frac {\widetilde{\nabla}_bF}{F}}.
\end{equation}
Quite remarkably, we thus find that the dynamics of an extended test body in the monopole approximation is independent of the body's mass. In case of a trivial coupling function $F$, equation (\ref{MonoF}) reproduces the well known general relativistic result. 

Interestingly, the mass of a body is not constant. Substituting (\ref{xiM}) into (\ref{MFdot}) we can solve the resulting differential equation to find explicitly the dependence of mass on the scalar function: $M = F^3M_0$ with $M_0=$const.

\section{Conclusions}\label{conclusion_sec}

\begin{table}
\caption{\label{tab_mag_eom_overview}Overview: MAG equations of motion.}
\begin{tabular}{ll}
\hline\hline
\multicolumn{2}{l}{{Lagrangian and conservation laws}}\\
\hline
&\\
\multicolumn{2}{l}{{${L} = {V}(g_{ij}, R_{ijk}{}^l, N_{ki}{}^j) + F(g_{ij},R_{kli}{}^j, N_{kl}{}^i)\,L_{\rm mat}(\psi^A, \nabla_i\psi^A)$}}\\
&\\
\multicolumn{2}{l}{{$\widetilde{\nabla}_j(F\Delta^i{}_k{}^j) =  F(\Sigma_k{}^i - t_k{}^i + N_{nm}{}^i\Delta^m{}_k{}^n - N_{nk}{}^m\Delta^i{}_m{}^n)$}}\\
&\\
\multicolumn{2}{l}{{$\widetilde{\nabla}_j\{F(\Sigma_k{}^j + \Delta^m{}_n{}^j N_{km}{}^n)\} = -\,F\Delta^m{}_n{}^i(\widetilde{R}_{kim}{}^n - \widetilde{\nabla}_kN_{im}{}^n) - L_{\rm mat}\widetilde{\nabla}_kF$}}\\
&\\
\hline
\multicolumn{2}{l}{{Equations of motion (any order)}}\\
\hline
&\\
See equations (\ref{int_eom_1_general}) and (\ref{int_eom_2_general}).&\\
&\\
\hline
\multicolumn{2}{l}{{Equations of motion (pole-dipole order)}}\\
\hline
&\\
\multicolumn{2}{l}{{${\frac {D{\cal P}^a}{ds}} = {\frac 12}\widetilde{R}^a{}_{bcd}v^b{\cal J}^{cd} + Fq^{cbd}\widetilde{\nabla}^aN_{dcb}-\, \xi\widetilde{\nabla}^aF - \xi^b\widetilde{\nabla}_b\widetilde{\nabla}^aF$}}\\
&\\
\multicolumn{2}{l}{{${\frac {D{\cal J}^{ab}}{ds}} = -\,2v^{[a}{\cal P}^{b]} + 2F(q^{cd[a}N^{b]}{}_{cd} + q^{c[a|d|}N_{dc}{}^{b]} + q^{[a|cd|}N_d{}^{b]}{}_c) - 2\xi^{[a}\widetilde{\nabla}^{b]}F$}}\\
&\\
$L^{ab} = 2p^{[ab]}$ & $S^{ab} = -2h^{[ab]}$ \\
${\cal P}^a = F(p^a + N^a{}_{cd}h^{cd}) + p^{ba}\widehat{\nabla}_bF$ & ${\cal J}^{ab}=F(L^{ab} + S^{ab})$\\
&\\
\hline
\multicolumn{2}{l}{{Equations of motion (monopole order)}}\\
\hline
&\\
$M{\frac {Dv^a}{ds}} = -\,\xi(g^{ab} - v^av^b){\frac {\widetilde{\nabla}_bF}F}$ & $M=p^a v_a $ \\
&\\
\hline\hline
\end{tabular}
\end{table}

\begin{table}
\caption{\label{tab_st_eom_overview}Overview: Scalar-tensor equations of motion (Einstein frame).}
\begin{tabular}{ll}
\hline\hline
\multicolumn{2}{l}{{Lagrangian and conservation law}}\\
\hline
&\\
\multicolumn{2}{l}{{$L = {\frac 1{2\kappa}}\left(- \widetilde{R} + g^{ij}\gamma_{AB}\partial_i\varphi^A\partial_j\varphi^B - 2U\right) + {\frac 1{F^4}}L_{\rm mat}(\psi,\partial\psi,F^{-2}g_{ij})$$\phantom{AAAA}$}}\\
&\\
\multicolumn{2}{l}{{$F = F(\varphi^A) \qquad {U} = {U}(\varphi^A) \qquad {\gamma}_{AB} = {\gamma}{}_{AB}(\varphi^A)$}}\\
&\\
\multicolumn{2}{l}{{$\widetilde{\nabla}^it_{ij} = {\frac 1F}\left(4t_{ij} - g_{ij}t\right)\partial^iF$}}\\
&\\
\hline
\multicolumn{2}{l}{{Equations of motion (any order)}}\\
\hline
&\\
See equation (\ref{int_eom_ST}).&\\
&\\
\hline
\multicolumn{2}{l}{{Equations of motion (pole-dipole order)}}\\
\hline
&\\
\multicolumn{2}{l}{{${\frac {D{\cal P}^a}{ds}} = {\frac 12}\widetilde{R}^a{}_{bcd}v^b{\cal J}^{cd} - \xi\widetilde{\nabla}^aF^{-4} - \xi^b\widetilde{\nabla}_b\widetilde{\nabla}^aF^{-4}$}}\\
&\\
\multicolumn{2}{l}{{${\frac {D{\cal J}^{ab}}{ds}} = -\,2v^{[a}{\cal P}^{b]} - 2\xi^{[a}\widetilde{\nabla}^{b]}F^{-4}$}}\\
&\\
$L^{ab} = 2p^{[ab]}$ &  \\
${\cal P}^a = F^{-4}p^a + p^{ba}\widetilde{\nabla}_bF^{-4}$ & ${\cal J}^{ab} = F^{-4}L^{ab}$\\
&\\
\hline
\multicolumn{2}{l}{{Equations of motion (monopole order)}}\\
\hline
&\\
${\frac {Dv^a}{ds}} = -\,(g^{ab} - v^av^b){\frac {\widetilde{\nabla}_bF}{F}}$ &  \\
&\\
\hline\hline
\end{tabular}
\end{table}

We have presented a general multipolar framework of covariant test body equations of motion for standard metric-affine gravity, as well as its extensions with nonminimal coupling between matter and gravity. Our results cover gauge theories of gravity (based on spacetime symmetry groups), and various so-called $f(R)$ models (and their generalizations), as well as scalar-tensor gravity. 

Our results unify and extend a whole set of works \cite{Hehl:1971,Bailey:Israel:1975,Stoeger:Yasskin:1979,Stoeger:Yasskin:1980,Puetzfeld:2007,Puetzfeld:Obukhov:2008:1,Puetzfeld:Obukhov:2008:2,Puetzfeld:Obukhov:2013,Hehl:Obukhov:Puetzfeld:2013,Puetzfeld:Obukhov:2013:3,Puetzfeld:Obukhov:2013:4,Puetzfeld:Obukhov:2014:1}.
In particular they can be viewed as a completion of the program initiated in \cite{Puetzfeld:2007}, in which a noncovariant Papapetrou \cite{Papapetrou:1951:3} type of approach was used. The general equations of motion (\ref{int_eom_1_general}) and (\ref{int_eom_2_general}) cover {\it all}\, of the previously reported cases. As demonstrated explicitly, the master equation (\ref{master}) allows for a quick adoption to any physical theory, as soon as the conservation laws and (multi-)current structure is fixed. Table \ref{tab_mag_eom_overview} contains an overview of our main results in the context of metric-affine gravity. In particular, the explicit equations of motion at the pole-dipole and monopole order are given. We stress that these equations hold for the most general case, i.e.\ including a general nonminimal coupling between matter and gravity. The results for standard (minimal) MAG are easily recovered by choosing a trivial coupling function $F$. 

Our analysis reveals how the new geometrical structures in generalized theories of gravity couple to matter, which in turn should be used for the design of experimental tests of gravity beyond the Einsteinian (purely Riemannian) geometrical picture. In the case of {\it minimal} coupling, we have once more confirmed -- now in a very general context -- that only matter matter with microstructure (such as the intrinsic hypermomentum, including spin, dilation and shear charges) allows for the detection of post-Riemannian structures. However, in gravitational theories with {\it nonminimal} coupling, there seems to be a loophole which may proof to be interesting for possible experiments; i.e.\ there is an {\it indirect} coupling of new geometrical quantities to regular matter via the nonminimal coupling function $F$. This may be exploited to devise new strategies to detect possible post-Riemannian spacetime features in future experiments. 

In addition to the results in MAG, we have explicitly worked out the test body equations of motion for a very general class of scalar-tensor gravitational theories. Table \ref{tab_st_eom_overview} contains an overview of our main results in the context of this theory class. Again the equations of motion at the pole-dipole and monopole order for a general coupling function $F$, which now depends on the scalar degrees of freedom, are explicitly given.  

We hope that our covariant unified multipolar framework sheds more light on the systematic test of gravitational theories by means of extended and microstructured test bodies. We would like to conclude with a statement by Einstein \cite{Einstein:1921} who stressed that 
\begin{quotation}
``[...] the question whether this continuum has a Euclidean, Riemannian, or any other structure is a question of physics proper which must be answered by experience, and not a question of a convention to be chosen on grounds of mere expediency.''
\end{quotation}

\section{Acknowledgements}
We would like to thank A.\ Trautman (University of Warsaw), W.G.\ Dixon (University of Cambridge), J.\ Madore (University of Paris South), and W.\ Tulczyjew (INFN Napoli) for sharing their insights into gravitational multipole formalisms and discussing their pioneering works with us. Furthermore, we would like to thank F.W.\ Hehl (University of Cologne) for fruitful discussion on gauge gravity models, in particular on Metric-Affine Gravity (MAG). D.P.\ was supported by the Deutsche Forschungsgemeinschaft (DFG) through the grants LA-905/8-1/2 and SFB 1128/1 (geo-Q).

\section*{Appendix}

\section*{A Conventions \& Symbols}\label{conventions_app}

\begin{table}
\caption{\label{tab_symbols1}Directory of symbols.}
\begin{tabular}{lp{8.5cm}}
\hline\hline
Symbol & Explanation\\
\hline
&\\
\hline
\multicolumn{2}{l}{{Geometrical quantities}}\\
\hline
$x^{a}$, $s$ & Coordinates, proper time \\
$g_{a b}$ & Metric\\
$\delta^a_b$ & Kronecker symbol \\
$g$ & Determinant of the metric \\
$\sigma$ & World-function\\
$\Gamma_{a b}{}^c$ & Connection \\
$\widetilde{\Gamma}_{a b}{}^c$ & Riemannian connection \\
$\overline{\Gamma}_{a b}{}^c$ & Transposed connection \\
$R_{a b c}{}^d$& Curvature \\
$Q_{a b c}$ & Nonmetricity \\
$Q_{a}$ & Weyl covector \\
$T_{a b}{}^c$ & Torsion \\
$K_{ab}{}^c$ & Contortion \\ 
$N_{a b}{}^c$ & Distortion \\
$R_{ab}, R$ & Ricci tensor / scalar \\ 
$(\sigma^A{}_B)_j{}^i$ & Generators of general coordinate transformations\\
$V, {\mathfrak V}$ & Gravitational Lagrangian (density) \\
$({H}^{\dots}{}_{\dots}, {\mathfrak H}^{\dots}{}_{\dots}), ({M}^{\dots},{\mathfrak M}^{\dots})$  & Generalized gravitational field momenta (densities)\\
${E}^{ab}{}_b, {\mathfrak E}^{ab}{}_b$ & Gravitational hypermomentum (density) \\
${E}_a{}^b, {\mathfrak E}_a{}^b$  & Generalized gravitational energy-momentum (density)\\
$I$ & General action\\
$L, {\mathfrak L}$ & Total Lagrangian (density)\\
$\Phi^J$ & General set of fields \\
$\zeta^a$ & (Generalized) Killing vector \\
$J^{Aj}, {\mathfrak J}^{Aj}$ & General dynamical currents (densities)\\
${\stackrel Jg}{}_{ij}$ & Jordan frame metric\\
${\stackrel {J}{L}},{\stackrel {J}{\mathfrak L}} $ & Total Lagrangian (density) in Jordan frame\\
\hline\hline
\end{tabular}
\end{table}

\begin{table}
\caption{\label{tab_symbols2}Directory of symbols.}
\begin{tabular}{lp{8.5cm}}
\hline\hline
Symbol & Explanation\\
\hline
&\\
\hline
\multicolumn{2}{l}{{Matter quantities}}\\
\hline
$\psi^A$ & General matter field \\
${L}_{\rm mat}, {\mathfrak L}_{\rm mat}$ & Matter Lagrangian (density) \\
${\Sigma}_a{}^b, {\mathfrak T}_a{}^b$ & Canonical energy-momentum (density) of matter\\
${\Delta}^a{}_b{}^c, {\mathfrak S}^a{}_b{}^c $ & Canonical hypermomentum (density) of matter\\
${t}_{ab},{\mathfrak t}_{ab}$ & Metrical energy-momentum (density) \\
$\tau_{ab}{}^c$ & Spin current\\
$\Delta^a$ & Dilation current\\
$v^a$ & Four velocity\\
$p$ & Pressure (hyperfluid) \\
$P_a$ & Momentum density (hyperfluid)\\
$J_a{}^b$ & Intrinsic hypermomentum density (hyperfluid)\\
$F$, $A$ & Coupling function\\
$F^{ab}$ & Variation of the coupling function \\
${\stackrel \zeta I}{}^a$ & Induced conserved current \\ 
$K^{\dot{A}}, {\mathfrak K}^{\dot{A}} $ & General material currents (densities)\\
$j^{\dots}, i^{\dots}, m^{\dots}, p^{\dots}, k^{\dots}, $& Integrated moments\\
$h^{\dots}, q^{\dots}, \mu^{\dots}, \xi^{\dots}$  & \\
$F^{ab}$ & Electromagnetic field \\
$J^{a}$ & Electric current \\
$L^{ab}, S^{ab}$ & Total orbital / spin angular momentum \\
${\cal P}^a, {\cal J}^{ab}$ & Generalized total momentum / angular momentum \\
$\Upsilon^{ab}$ & Total hypermomentum \\
$M$ & Generalized testbody mass \\
$D$ & Intrinsic dilation moment \\
$\varphi^A$ & Multiplet of scalar fields \\
$\kappa$ & Einstein's gravitational coupling constant \\ 
\hline\hline
\end{tabular}
\end{table}

\begin{table}
\caption{\label{tab_symbols3}Directory of symbols.}
\begin{tabular}{lp{8.5cm}}
\hline\hline
Symbol & Explanation\\
\hline
&\\
\hline
\multicolumn{2}{l}{{Auxiliary quantities}}\\
\hline
$\Omega_a, \Omega_a{}^b, \Omega_a{}^{bc}, \Omega_a{}^{bcd}, \Omega'{}_k$ & Auxiliary variables (Noether identities) \\
$\rho^{abc}{}_d, \mu^{ab}{}_c $ & Partial derivatives of the total Lagrangian \\
${\mathfrak L}_0$ & Auxiliary Lagrangian density \\
${\stackrel 0 \rho}{}^{abc}{}_d, {\stackrel 0 \mu}{}^{ab}{}_c $ & Partial derivatives of the coupling function \\
$A,B, \dots; \dot{A}, \dot{B}, \dots$ & General multi-indices \\
$\Lambda_{jB}{}^A, \Pi^A{}_{\dot{B}}$ & General functions of external classical fields\\
$U_{abc}{}^d$, $V_{ab}{}^c$, $\kappa^{abc}$ & Auxiliary variables (MAG equations of motion) \\
$A_k$ & Derivative of the coupling function\\
$\alpha^{y_0}{}_{y_1 \dots y_n}$, $\beta^{y_0}{}_{y_1 \dots y_n}$, $\gamma^{y_0}{}_{y_1 \dots y_n}$& Expansion coefficients of the parallel propagator\\
${\stackrel {d}{q}}{}^{abc}, {\stackrel {s}{q}}{}^{abc}$ & Decomposition pieces of the q-moments\\
$Z^a$ & Trace of the ${\stackrel {d}{q}}$ moment \\
${\gamma}_{AB}, {\stackrel J\gamma}_{AB}$ & General function of scalar fields (in Jordan frame)\\
${U}, {\stackrel JU}$ & General potential of scalar fields (in Jordan frame)\\
$\Xi^{ab}$ & Auxiliary variable (Einstein frame)\\
$\Phi^{y_1 \dots y_n y_ 0}{}_{x_0}$, $\Psi^{y_2 \dots y_n+1 y_ 0 y_1}{}_{x_0 x_1}$ &  Auxiliary variables multipole expansions \\
\hline
\multicolumn{2}{l}{{Operators / accents}}\\
\hline
$\partial_a$, ${\cal L}_\zeta$ & (Partial, Lie) derivative \\ 
$\nabla_a$ & Covariant derivative \\ 
$\widetilde{\nabla}_a$, ``${\phantom{A}}_{;a}$'' & Riemannian covariant derivative \\ 
$\overline{\nabla}_a$ & Transposed covariant derivative \\ 
$\widehat{\nabla}{}_a$ & Covariant density derivative \\ 
$\check{\nabla}{}_a$ & Riemannian covariant density derivative \\ 
${\stackrel * \nabla}{}_a$ & Modified covariant density derivative \\
$\frac{D}{ds} = $``$\dot{\phantom{a}}$'' & Total derivative \\
$\delta, \subsvar $ & Variation, substantial variation \\
${\mathfrak B}_{\dots}{}^{\dots}$ & (Bi-)Tensor density \\ 
$g^{y_0}{}_{x_0}$, $G^Y{}_X$ & (Generalized) parallel propagator\\
``$\stackrel{J}{\phantom{AA}}$'' & Jordan frame quantity\\
``$\widetilde{\phantom{AA}}$'' & Riemannian quantity\\
``$[ \dots ]$''& Coincidence limit\\
\hline\hline
\end{tabular}
\end{table}
 
In the following we summarize our conventions, and collect some frequently used formulas. A directory of symbols used throughout the text can be found in tables \ref{tab_symbols1}, \ref{tab_symbols2}, \ref{tab_symbols3}.

For an arbitrary $k$-tensor $T_{a_1 \dots a_k}$, the symmetrization and antisymmetrization are defined by
\begin{eqnarray}
T_{(a_1\dots a_k)} &:=& {\frac 1{k!}}\sum_{I=1}^{k!}T_{\pi_I\!\{a_1\dots a_k\}},\label{S}\\
T_{[a_1\dots a_k]} &:=& {\frac 1{k!}}\sum_{I=1}^{k!}(-1)^{|\pi_I|}T_{\pi_I\!\{a_1\dots a_k\}},\label{A}
\end{eqnarray}
where the sum is taken over all possible permutations (symbolically denoted by $\pi_I\!\{a_1\dots a_k\}$) of its $k$ indices. As is well-known, the number of such permutations is equal to $k!$. The sign factor depends on whether a permutation is even ($|\pi| = 0$) or odd ($|\pi| = 1$). The number of independent components of the totally symmetric tensor $T_{(a_1\dots a_k)}$ of rank $k$ in $n$ dimensions is equal to the binomial coefficient ${{n-1+k}\choose{k}} = (n-1+k)!/[k!(n-1)!]$, whereas the number of independent components of the totally antisymmetric tensor $T_{[a_1\dots a_k]}$ of rank $k$ in $n$ dimensions is equal to the binomial coefficient ${{n}\choose{k}} = n!/[k!(n-k)!]$. For example, for a second rank tensor $T_{ab}$ the symmetrization yields a tensor $T_{(ab)} = {\frac 12}(T_{ab} + T_{ba})$ with 10 independent components, and the antisymmetrization yields another tensor $T_{[ab]} = {\frac 12}(T_{ab} - T_{ba})$ with 6 independent components.
 
In the derivation of the equations of motion we made use of the bitensor formalism, see, e.g., \cite{Synge:1960,DeWitt:Brehme:1960,Poisson:etal:2011} for introductions and references. In particular, the world-function is defined as an integral $\sigma(x,y) := \frac{1}{2} \epsilon \left( \int\limits_x^y d\tau \right)^2$ over the geodesic curve connecting the spacetime points $x$ and $y$, where $\epsilon = \pm 1$ for timelike/spacelike curves. Note that our curvature conventions differ from those in \cite{Synge:1960,Poisson:etal:2011}. Indices attached to the world-function always denote covariant derivatives, at the given point, i.e.\ $\sigma_y:= \widetilde{\nabla}_y \sigma$, hence we do not make explicit use of the semicolon in case of the world-function. The parallel propagator by $g^y{}_x(x,y)$ allows for the parallel transportation of objects along the unique geodesic that links the points $x$ and $y$. For example, given a vector $V^x$ at $x$, the corresponding vector at $y$ is obtained by means of the parallel transport along the geodesic curve as $V^y = g^y{}_x(x,y)V^x$. For more details see, e.g., \cite{Synge:1960,DeWitt:Brehme:1960} or section 5 in \cite{Poisson:etal:2011}. A compact summary of useful formulas in the context of the bitensor formalism can also be found in the appendices A and B of \cite{Puetzfeld:Obukhov:2013}.

We start by stating, without proof, the following useful rule for a bitensor $B$ with arbitrary indices at different points (here just denoted by dots):
\begin{eqnarray}
\left[B_{\dots} \right]_{;y} = \left[B_{\dots ; y} \right] + \left[B_{\dots ; x} \right]. \label{synges_rule}
\end{eqnarray}
Here a coincidence limit of a bitensor $B_{\dots}(x,y)$ is a tensor 
\begin{eqnarray}
\left[B_{\dots} \right] = \lim\limits_{x\rightarrow y}\,B_{\dots}(x,y),\label{coin}
\end{eqnarray}
determined at $y$. Furthermore, we collect the following useful identities: 
\begin{eqnarray}
&&\sigma_{y_0 y_1 x_0 y_2 x_1} = \sigma_{y_0 y_1 y_2 x_0 x_1} = \sigma_{x_0 x_1 y_0 y_1 y_2 }, \label{rule_1} \\
&&g^{x_1 x_2} \sigma_{x_1} \sigma_{x_2} = 2 \sigma = g^{y_1 y_2} \sigma_{y_1} \sigma_{y_2}, \label{rule_2}\\
&&\left[ \sigma \right]=0, \quad  \left[ \sigma_x \right] = \left[ \sigma_y \right]  = 0, \label{rule_3} \\
&& \left[ \sigma_{x_1 x_2} \right] =  \left[ \sigma_{y_1 y_2} \right] = g_{y_1 y_2}, \quad
\left[ \sigma_{x_1 y_2} \right] =  \left[ \sigma_{y_1 x_2} \right] = - g_{y_1 y_2}, \label{rule_5}\\ 
&& \left[ \sigma_{x_1 x_2 x_3} \right] = \left[ \sigma_{x_1 x_2 y_3} \right] = \left[ \sigma_{x_1 y_2 y_3} \right] = \left[ \sigma_{y_1 y_2 y_3} \right] = 0,\label{rule_6}\\
&&\left[g^{x_0}{}_{y_1} \right] = \delta^{y_0}{}_{y_1}, \quad \left[g^{x_0}{}_{y_1 ; x_2} \right] = \left[g^{x_0}{}_{y_1 ; y_2} \right] = 0, \label{rule_7} \\
&& \left[g^{x_0}{}_{y_1 ; x_2 x_3} \right] = \frac{1}{2} \widetilde{R}{}^{y_0}{}_{y_1 y_2 y_3}. \label{rule_8}
\end{eqnarray}

\section*{B Covariant expansions}\label{expansion_app}

Here we briefly summarize the covariant expansions of the second derivative of the world-function, and the derivative of the parallel propagator:
\begin{eqnarray}
\sigma^{y_0}{}_{x_1} &=& g^{y'}{}_{x_1}\biggl( -\,\delta^{y_0}{}_{y'} +\,\sum\limits_{k=2}^\infty\,{\frac {1}{k!}}\,\alpha^{y_0}{}_{y'y_2\!\dots \!y_{k+1}}\sigma^{y_2}\cdots\sigma^{y_{k+1}}\biggr)\!,\label{app_expansion_1}\\
\sigma^{y_0}{}_{y_1} &=& \delta^{y_0}{}_{y_1} -\,\sum\limits_{k=2}^\infty\,{\frac {1}{k!}}\,\beta^{y_0}{}_{y_1y_2\dots y_{k+1}} \sigma^{y_2}\!\cdots\!\sigma^{y_{k+1}}, \label{app_expansion_2} \\
g^{y_0}{}_{x_1 ; x_2} &=& g^{y'}{\!}_{x_1} g^{y''}{\!}_{x_2}\biggl({\frac 12} 
\widetilde{R}{}^{y_0}{}_{y'y''y_3}\sigma^{y_3}\!+\!\sum\limits_{k=2}^\infty\,{\frac {1}{k!}}\,\gamma^{y_0}{}_{y'y''y_3\dots y_{k+2}}\sigma^{y_3}\!\cdots\!\sigma^{y_{k+2}}\!\biggr)\!,\nonumber \\  \label{app_expansion_3} \\
g^{y_0}{}_{x_1 ; y_2} &=& g^{y'}{\!}_{x_1} \biggl({\frac 12} \widetilde{R}{}^{y_0}{}_{y'y_2y_3}\sigma^{y_3}\!+\!\sum\limits_{k=2}^\infty\,{\frac {1}{k!}}\,\gamma^{y_0}{}_{y'y_2y_3\dots y_{k+2}}\sigma^{y_3}\!\cdots\!\sigma^{y_{k+2}}\!\biggr).\label{app_expansion_4}\\
G^{Y_0}{}_{X_1 ; x_2} &=& G^{Y'}{\!}_{X_1} g^{y''}{\!}_{x_2} \sum\limits_{k=1}^\infty\,{\frac {1}{k!}}\,\gamma^{Y_0}{}_{Y'y''y_3\dots y_{k+2}}\sigma^{y_3}\!\cdots\!\sigma^{y_{k+2}}, \label{app_expansion_5} \\
G^{Y_0}{}_{X_1 ; y_2} &=& G^{Y'}{\!}_{X_1} \sum\limits_{k=1}^\infty\,{\frac {1}{k!}}\,\gamma^{Y_0}{}_{Y'y_2y_3\dots y_{k+2}}\sigma^{y_3}\!\cdots\!\sigma^{y_{k+2}}.\label{app_expansion_6}
\end{eqnarray}
The coefficients $\alpha, \beta, \gamma$ in these expansions are polynomials constructed from the Riemann curvature tensor and its covariant derivatives. The first coefficients read as follows:
\begin{eqnarray}
\alpha^{y_0}{}_{y_1y_2y_3} &=& - \frac{1}{3} \widetilde{R}{}^{y_0}{}_{(y_2y_3)y_1},\label{a1}\\
\beta^{y_0}{}_{y_1y_2y_3} &=& \frac{2}{3}\widetilde{R}{}^{y_0}{}_{(y_2y_3)y_1},\label{be1}\\
\alpha^{y_0}{}_{y_1y_2y_3y_4} &=& \frac{1}{2} \widetilde{\nabla}_{(y_2}\widetilde{R}{}^{y_0}{}_{y_3y_4)y_1},\label{al2}\\
\beta^{y_0}{}_{y_1y_2y_3y_4} &=& - \frac{1}{2} \widetilde{\nabla}_{(y_2} \widetilde{R}{}^{y_0}{}_{y_3y_4)y_1},\label{be2}\\
\nonumber\\
\gamma^{y_0}{}_{y_1y_2y_3y_4}&=& \frac{1}{3} \widetilde{\nabla}_{(y_3} \widetilde{R}{}^{y_0}{}_{|y_1|y_4)y_2}.\label{ga}
\end{eqnarray}
In addition, we also need the covariant expansion of a vector:
\begin{eqnarray} 
A_x = g^{y_0}{}_x\,\sum\limits_{k=0}^\infty\,{\frac {(-1)^k}{k!}} \, A_{y_0;y_1\dots y_k}\,\sigma^{y_1}\cdots\sigma^{y_k}.\label{Ax}
\end{eqnarray}

\section*{C Explicit form}\label{explicit_app}

Here we make contact with our notation in \cite{Puetzfeld:Obukhov:2013:3} to facilitate a direct comparison to the results in there. 

We introduce the auxiliary variables 
\begin{eqnarray}
\Phi^{y_1\dots y_ny_0}{}_{x_0} &:=& \sigma^{y_1} \cdots \sigma^{y_n} g^{y_0}{}_{x_0},\label{Phi}\\
\Psi^{y_1\dots y_ny_0y'}{}_{x_0x'} &:=& \sigma^{y_1} \cdots \sigma^{y_n} g^{y_0}{}_{x_0}g^{y'}{}_{x'}.
\label{Psi}
\end{eqnarray}
Their derivatives 
\begin{eqnarray}
\Psi^{y_1\dots y_ny_0y'}{}_{x_0x';z} &=& \sum^{n}_{a=1}\sigma^{y_1}\cdots\sigma^{y_a}{}_z\cdots\sigma^{y_n}g^{y_0}{}_{x_0}g^{y'}{}_{x'} \nonumber \\
&&+ \sigma^{y_1} \cdots \sigma^{y_n}\left(g^{y_0}{}_{x_0;z}g^{y'}{}_{x'} + g^{y_0}{}_{x_0}g^{y'}{}_{x';z}\right)\label{dPsi},\\
\Phi^{y_1\dots y_ny_0}{}_{x_0;z} &=& \sum^{n}_{a=1}\sigma^{y_1}\cdots\sigma^{y_a}{}_z\cdots\sigma^{y_n}g^{y_0}{}_{x_0}\nonumber \\
&&+ \sigma^{y_1} \cdots \sigma^{y_n}\,g^{y_0}{}_{x_0;z},\label{dPhi}
\end{eqnarray}
can be straightforwardly evaluated by using the expansions from the previous appendix. 

In terms of (\ref{Phi}) and (\ref{Psi}) the integrated conservation laws (\ref{cons1f}) and (\ref{cons2f}) take the form: 
\begin{eqnarray}
&&{\frac{D}{ds}} \int \Psi^{y_1\dots y_ny_0y'}{}_{x_0x'}{\mathfrak S}^{x_0 x' x_2} d\Sigma_{x_2} = \nonumber \\
&&
\int \Psi^{y_1\dots y_ny_0y'}{}_{x_0x'}\Big( - U_{x''''}{}^{x'}{}_{x''}{}^{x_0}{}_{x'''}{\mathfrak S}^{x'' x''' x''''} 
+ {\mathfrak T}^{x' x_0} - {\mathfrak t}^{x' x_0}\Big) w^{x_2} d\Sigma_{x_2} \nonumber \\
&&+\,\int\Psi^{y_1\dots y_ny_0y'}{}_{x_0x';x''}{\mathfrak S}^{x_0 x' x''}w^{x_2} d\Sigma_{x_2}\nonumber \\
&&+ \int v^{y_{n+1}}\Psi^{y_1\dots y_ny_0y'}{}_{x_0x';y_{n+1}}{\mathfrak S}^{x_0 x' x_2}d\Sigma_{x_2}, \label{int_eom_1}
\end{eqnarray}
\begin{eqnarray}
&&{\frac{D}{ds}} \int\Phi^{y_1\dots y_ny_0}{}_{x_0}{\mathfrak T}^{x_0 x_2} d\Sigma_{x_2} = \int \Phi^{y_1\dots y_ny_0}{}_{x_0} \Big(-V_{x''}{}^{x_0}{}_{x'}{\mathfrak T}^{x' x''} \nonumber \\
&&- R^{x_0}{}_{x''' x' x''}{\mathfrak S}^{x' x'' x'''} - \frac{1}{2} Q^{x_0}{}_{x'' x'}{\mathfrak t}^{x' x''} -\,A^{x_0}{\mathfrak L}_{\rm mat}\Big) w^{x_2} d\Sigma_{x_2} \nonumber\\
&&+ \int\Phi^{y_1\dots y_ny_0}{}_{x_0;x'}{\mathfrak T}^{x_0 x'}w^{x_2} d\Sigma_{x_2} 
+ \int v^{y_{n+1}}\Phi^{y_1\dots y_ny_0}{}_{x_0;y_{n+1}}{\mathfrak T}^{x_0 x_2} d\Sigma_{x_2}. \nonumber \\\label{int_eom_2}
\end{eqnarray}
This form allows for a direct comparison to (29) and (30) in \cite{Puetzfeld:Obukhov:2013:3}. Explicitly, in terms of (\ref{Phi}) and (\ref{Psi}) the integrated moments from (\ref{jMAG})--(\ref{mMAG}) are given by:
\begin{eqnarray}
p^{y_1\dots y_n y_0} \!\!&:=&\!\! (-1)^n\!\!\int\limits_{\Sigma(\tau)}\!\!\Phi^{y_1\dots y_n y_0}{}_{x_0}{\mathfrak T}^{x_0 x_1}d\Sigma_{x_1},\\
k^{y_2\dots y_{n+1} y_0 y_1} \!\!&:=&\!\! (-1)^n\!\!\int\limits_{\Sigma(\tau)}\!\!\Psi^{{y_2}\dots{y_{n+1} y_0 y_1}}{}_{x_0 x_1}{\mathfrak T}^{x_0 x_1}w^{x_2}d\Sigma_{x_2},\\
h^{y_2\dots y_{n+1}y_0 y_1} \!\!&:=&\!\! (-1)^n\!\!\int\limits_{\Sigma(\tau)}\!\!\Psi^{y_2 \dots y_{n+1}y_0 y_1}{}_{x_0 x_1 }{\mathfrak S}^{x_0 x_1 x_2}d\Sigma_{x_2},\label{Smom}\\
q^{y_3\dots y_{n+2}y_0 y_1 y_2} \!\!&:=&\!\! (-1)^n\!\!\int\limits_{\Sigma(\tau)}\!\!\Psi^{y_3 \dots y_{n+2} y_0 y_1}{}_{x_0 x_1} g^{y_2}{}_{x_2}{\mathfrak S}^{x_0 x_1 x_2 }w^{x_3}d\Sigma_{x_3},\\
\mu^{y_2\dots y_{n+1} y_0 y_1} \!\!&:=&\!\! (-1)^n\!\!\int\limits_{\Sigma(\tau)}\!\!\Psi^{y_2 \dots y_{n+1} y_0 y_1}{}_{x_0 x_1}{\mathfrak t}^{x_0 x_1}w^{x_2}d\Sigma_{x_2},\\
\xi^{y_1\dots y_{n}} \!\!&:=&\!\! (-1)^n\!\!\int\limits_{\Sigma(\tau)}\!\!\sigma^{y_1}\cdots\sigma^{y_{n}}{\mathfrak L}_{\rm mat}w^{x_2}d\Sigma_{x_2}.
\end{eqnarray}

\bibliographystyle{unsrt}

\bibliography{puetzfeld_obukhov_eom_proceedings_2013}

\end{document}